\documentclass[prl, aps, 12pt, showpacs,epsfig]{revtex4}
\usepackage{graphicx}
\usepackage{amsmath}
\begin{document}

%\title{Phenomenology of Lepton Triplet }
\title{Triplet with doubly-changed lepton at the LHC}

\author{Teng Ma, Bin Zhang}
\email{zb@mail.tsinghua.edu.cn (Communication author)}
 \affiliation{ Department of Physics,
Tsinghua University, Beijing, 100084, China\\
Center for
High Energy Physics, Tsinghua University, Beijing, 100084, China}
\author{Giacomo~Cacciapaglia}
\affiliation{Universit\'e de Lyon, F-69622 Lyon, France; Universit\'e Lyon 1, Villeurbanne;
CNRS/IN2P3, UMR5822, Institut de Physique Nucl\'aire de Lyon,
F-69622 Villeurbanne Cedex, France}

\begin{abstract}
We studied the signatures of an electroweak SU(2) lepton triplet which contains a doubly charged heavy lepton. The most interesting and easily detected final states are originating from the doubly charged lepton pair and single production, that we analyzed at a center of mass energy of 14 TeV. We propose useful kinematic cuts to reduce the backgrounds and computed the integrated luminosity needed to discover or exclude the doubly charged leptons for various masses.

\end{abstract}


 \maketitle

\section{Introduction}

Although the Standard Model (SM) of electroweak interactions explains with great accuracy almost all of the available experimental data, there are good reasons to believe that it is not the ultimate theory to describe Nature.
The main motivation behind the expectation of new physics around the TeV scale is the naturalness of the Higgs mass, and consequently of the the electroweak scale.
The discovery of a light scalar with the properties expected for the SM Higgs boson at the LHC, announced on july 4$^{th}$ 2012, has completed the list of particles predicted by the SM, and also brought to reality the problem of naturalness.
In fact we now know that there exists a light scalar associated with the breaking of the electroweak scale, and a reason for its lightness is still lacking.

Many models of new physics addressing the naturalness problem have been proposed in the last few years, and a common feature is the prediction of the existence of new particles not far from the TeV scale, which is now being tested at the LHC.
Even though all the searches have given negative results, there is still plenty of space for the presence of new particles below the TeV. In fact, a hasty glimpse at the bounds on new physics, both supersymmetric and exotic, can leave the impression that most of the bounds are well above the TeV scale: the reason for this is that the quoted bounds refer to particles which are easy to detect and with large production rates.
Such cases often correspond to less motivated or interesting scenarios.
Light states are still allowed if their couplings are suppressed, or they decay into finals states affected by large backgrounds, or they are not efficiently produced at the LHC.
In this paper we are interested in the latter case: we will in fact consider the LHC phenomenology of heavy exotic leptons, which have small production cross sections compared to colored states.
Such states would be interesting per se as a challenging channel for the LHC experiments, moreover some models predict their presence.
For instance, Little Higgs models, models of composite Higgs and extra dimensions all include leptonic partners together with quark partners.
If partners of the top are necessary to cancel the large quadratic divergences generated by the top mass~\cite{Schmaltz:2005ky}, leptonic partners are necessary for a consistent inclusion of the lepton mass generation and a study of their phenomenology in models with T-parity can be found in~\cite{Cacciapaglia:2009cv}.
New leptonic (un-colored) states are also necessary in some technicolor models~\cite{Frandsen:2009fs}, where they are introduced to cancel the gauge anomalies in the techni-gauge sector.
Another example of models where leptonic multiplets are introduced are models of see-saw that generate neutrino masses: the scenario of TeV scale masses have been considered, for instance, in~\cite{Franceschini:2008pz,delAguila:2008cj,delAguila:2008hw,delAguila:2008pw,Arhrib:2009mz}.

In most of the models, the new fermions have a mass which is not generated by the Brout-Englert-Higgs mechanism, therefore they are vector-like states in the sense that their left and right-handed projections have the same quantum numbers under the SM gauge group, as opposed to the chiral SM fermions.
The case of vector-like quarks~\cite{delAguila:2000aa,delAguila:2000rc,AguilarSaavedra:2009es,Cacciapaglia:2010vn,Buchkremer:2013bha,Aguilar-Saavedra:2013qpa} and partners of the top quark~\cite{AguilarSaavedra:2009es,DeSimone:2012fs} have been extensively studies in the literature, mostly due to their large production rates at the LHC.
As in the quark sector case, the vector-like leptons can decay to SM leptons only if Yukawa couplings with them is allowed, thus only a handful of possibilities for the gauge quantum numbers is allowed~\cite{delAguila:2008pw}.
The case where no Yukawa couplings are allowed is also of interest, as it may provide a Dark Matter candidate~\cite{Cirelli:2005uq}.
In Ref.~\cite{Delgado:2011iz} a thoroughly description of the case of a triplet with hypercharge $-1$ can be found, A. Delgado {\it et al} studied the phenomenology of these exotic leptons  in both low-energy  experiments and at the LHC. The main interest of the triplet is that it contains an exotic lepton with charge $-2$, which can lead to a clean same-sign lepton pair. The most interesting doubly-changed lepton pair production was studied in Ref.~\cite{Delgado:2011iz}. It is found that the exotic lepton can be discovered at the 7 TeV LHC if the mass less than 400 GeV, and can be discovered at the 14 TeV LHC for a 800 GeV mass.
A doubly charged lepton also appears in the case of a doublet with exotic hypercharge $Y = -3/2$.
The phenomenology of the doubly charged lepton from a triplet and doublet has been recently studied in~\cite{Alloul:2013raa} in a slightly different context, where the decays are due either to four-lepton interactions or mixing with the tau lepton uniquely.
In this work, we will reconsider the case of the triplet with a doubly-changed lepton, with couplings to light leptons. We will extend the  LHC phenomenology study in Ref.~\cite{Delgado:2011iz}, and include some additional signal channels and the even more sensitive lepton flavor violating channel in some cases.
The main novelty in our study is that we include additional irreducible contributions to the optimal tri-lepton channel coming from associated production of the doubly-charged lepton with the singly-charged one: this new contribution can significantly enhance the signal and the reach of the LHC.
We also include the study of decays of the singly-charged lepton: although these channels are less significant than the tri-lepton one, they are essential in determining the nature of the new lepton as they can allow a reconstruction of the multiplet the doubly-charged lepton belongs to.
Finally, we also consider the reach of a search based on lepton flavour violating channels, which may give stronger bounds than the tri-lepton case if the new leptons couple to both electrons and muons.
Our work mainly focuses on the possibility of dedicated search strategies at the high energy run that will start in 2015.

\section{the model}

In the following, we will consider an effective model consisting of the SM complemented by a vector-like $SU(2)$ triplet of exotic leptons with hypercharge $Y= - 1$:
\begin{equation}
	X = \begin{pmatrix} X^{0}\\X^{-}\\X^{--} \end{pmatrix} \in (3, -1).
\end{equation}
%where the subscripts $L,R$ refer to the chirality of the fermions.
The general Yukawa terms that give rise to the lepton masses without breaking gauge invariance or lepton number is, following the notation in~\cite{Delgado:2011iz},
\begin{eqnarray}
 -{\cal L}_Y &=& \lambda^{ij}_1\, \overline{L}^i_L H {e^j_R} + \lambda_2^{ij} \,\overline{L}^i_L H^c {\nu^j _R} + \lambda_3^j \, \overline{X_R} H^c L^j_L + M_1\,  \overline{X}_L X_R + h.c.\,,
\end{eqnarray}
where the subscript $L$ and $R$ refer to the chirality of the fermions, the indices $i,j = 1,2,3$ span over the 3 standard generations of leptons, $M_1$ is the gauge-invariant vector-like mass of the triplet, and Dirac masses for neutrinos are introduced.
The origin and nature of neutrino masses will be irrelevant for the phenomenology we are interested in the following, so we will not discuss any further this sector.
After the Brout-Englert-Higgs field $H$ develops its vacuum expectation value, the Yukawa terms will generate a mass mixing of the singly-charged and neutral component of $X$ with the standard leptons.
To simplify the discussion, we can use the flavor symmetry in the SM sector to diagonalise the standard Yukawa matrices $\lambda_1$ and $\lambda_2$, thus redefining the new Yukawa $\lambda_3$ accordingly.
In this basis, the mass terms look like:
\begin{eqnarray}
 - {\cal L}_{\rm mass} &=& m^i_1\, \overline{e}^i_L {e^i_R} + m_2^i\, \overline{\nu}^i_L {\nu^i _R} + \frac{1}{\sqrt{2}} {m_3^\ast}^i \, \overline{e}^i_L X^-_R + {m_3^\ast}^i\,  \overline{\nu}^i_L X^0_R + \nonumber \\
 & & M_1\, (\overline{X}^{--}_L X^{--}_R + \overline{X}^{-}_L X^{-}_R + \overline{X}^{0}_L X^{0}_R) + h.c.\,,
\end{eqnarray}
where the flavor index runs over $i=e, \mu, \tau$.
In the following we will closely follow the notation used in~\cite{Delgado:2011iz}: the mass matrices in the singly-charged and neutral sectors can be diagonalised by two unitary matrices each, $S_{E,N}$ for the left-handed and $T_{E,N}$ for the right-handed fields.
The mixing angles in the right-handed sector are suppressed compared to the right-handed ones by the ratio of the light fermion masses versus the mass $M_1$, therefore their contribution to the phenomenology can be neglected.
Furthermore, the couplings of the new states to the SM ones only depends on two vectors in flavor:
\begin{equation}
v_E^i = S_E^{\ast, 4 i}\,, \qquad \mbox{and} \quad v_N^i = S_N^{\ast,4 i}\,.
\end{equation}
Interesting relations between the couplings and masses have already been derived in~\cite{Delgado:2011iz}, so in this section we will limit ourselves to study in more detail some interesting limits in order to clarify important properties of the new leptons.

Firstly, the masses of the SM leptons are typically much smaller than the mass $M_1$, so we can start by neglecting their effect, i.e. setting $m_1^i = m_2^i = 0$.
In this case, the mass matrices can be diagonalised exactly, and we obtain for the mass eigenstates
\begin{equation}
M_{X^-}^2 = M_2^2 =  M_1^2 + \frac{1}{2} \sum_{i} |m_3^i|^2\,, \quad M_{X^0}^2 = M_3^2 = M_1^2 + \sum_{i} |m_3^i|^2\,, \quad \mbox{and} \quad M_{X^{--}}^2 = M_1^2\,;
\end{equation}
while the SM leptons remain massless.
From the above relations we see that a splitting between the triplet fields is induced by the new Yukawa couplings, and it is not proportional to the SM lepton masses.
Furthermore, we find the relations
\begin{equation}
M_{X^0} > M_{X^-} > M_{X^{--}} \quad \mbox{and} \quad M_{X^0}^2 - M_{X^{--}}^2 = 2 (M_{X^-}^2 - M_{X^{--}}^2)\,.
\end{equation}
The mixing angles are given by
\begin{equation}
v_E^i = - \frac{{m_3^*}^i}{\sqrt{2} M_2}\,, \quad v_N^i =  - \frac{{m_3^*}^i}{M_3} = \sqrt{2} \frac{M_2}{M_3} v_E^i\,.
\end{equation}
Note that no mixing in the right-handed sector is generated, thus showing that the right-handed mixing angles must be suppressed by the SM lepton masses.
The above flavor vectors will describe the couplings of the heavy leptons to the SM ones via gauge bosons ($W^\pm$, $Z$ and Higgs): for later convenience, we will introduce the quantity
\begin{equation}
\lambda = \sum_i |v_E^i|^2 = \frac{1}{2 M_2^2} \sum_i |m_3^i|^2
\end{equation}
which encodes the overall coupling strength of the mixing to standard leptons.
The mass eigenvalues can therefore be written in terms of $\lambda$ as
\begin{equation}
M_2^2 = \frac{1}{1-\lambda}  M_1^2\,, \quad M_3^2 = \frac{1+\lambda}{1-\lambda} M_1^2\,.
\end{equation}

Out of the lepton masses, only the tau one is considerably larger being at the GeV scale, therefore it is interesting to add in its effect and study how the masses and mixing terms are affected.
We will now turn on a non-zero $m_1^\tau$ (and keep to zero the masses of electon, muon and neutrinos).
The mass matrix in the charged sector can be diagonalised exactly.
Using the arbitrary phase of the triplet $X$ we can always chose $m_3^\tau$ to be real, and define a mixing angle~\cite{Cacciapaglia:2011fx}
\begin{equation}
\sin \theta_L = \frac{M_1 m_3^\tau}{\sqrt{2 (M_1^2 - m_\tau^2)^2 + M_1^2 (m_3^\tau)^2}}\,,
\end{equation}
where the physical tau mass $m_\tau$ is defined by the equation
\begin{equation}
(m_1^\tau)^2 = m_\tau^2 \left( 1 + \frac{(m_3^\tau)^2}{2 (M_1^2 - m_\tau^2)}\right) \,.
\end{equation}
The previous equation shows that the physical tau mass vanishes for $m_1^\tau = 0$.
The heavy lepton mass is now given by
\begin{equation}
M_{X^-}^2 = M_1^2 + \frac{1}{2} \left( \frac{M_1^2 (m_3^\tau)^2}{M_1^2 - m_\tau^2} + |m_3^e|^2 + |m_3^\mu|^2 \right)\,,
\end{equation}
with flavor mixings
\begin{equation}
v_E^e =  - \frac{\cos \theta_L}{\sqrt{2}} \frac{{m_3^*}^e}{M_{X^-}}\,, \quad
v_E^\mu =  - \frac{\cos \theta_L}{\sqrt{2}} \frac{{m_3^*}^\mu}{M_{X^-}}\,, \quad
v_E^\tau =  - \sin \theta_L\,.
\end{equation}
Note that a mixing angle of the tau in the right handed sector is now induced, with angles given by
\begin{equation}
\sin \theta_R = \frac{m_\tau}{M_1} \sin \theta_L\,.
\end{equation}
By expanding the above relations in $1/M_1$, we can show that the effect of the tau mass only appears at subleading order $\mathcal{O} (M_1^{-4})$, therefore it is safe to neglect its effect even when the new Yukawa couplings are small.

Before discussing the phenomenology of the model, it is important to assess how large the coupling to the light fermions can be.
It is well known that the mixing with a heavy lepton will modify the coupling of the light ones to the electroweak bosons, and also induce flavor changing couplings of the $Z$ boson.
We can therefore impose two kind of bounds: the ones originating from flavor changing processes will typically involve the product of the mixing terms to two different generations as they arise from processes involving the transition of one flavor into another.
The modification to the coupling of a given family to the $Z$ on the other hand will be sensitive to a single parameter, therefore they are essential to extract an absolute bound on the parameters.
Constraints of the former type have been extensively studied in~\cite{Delgado:2011iz}, so we will limit ourselves to recall the results.
The strongest constraints come from lepton flavor violation decays in the form $l_2 \to l_1 l^+ l^-$ mediated at tree level by the $Z$ boson, and they give the following bounds:
\begin{equation}
|v_E^e| |v_E^\mu| < 2.2 \cdot 10^{-6}\,, \quad |v_E^e| |v_E^\tau| < 8.6 \cdot 10^{-4}\,, \quad |v_E^\mu| |v_E^\tau| < 7.0 \cdot 10^{-4}\,. \label{eq:bounds1}
\end{equation}
Such bounds can be easily evaded by setting two of the three mixings to be very small.
Thus, to assess the maximum allowed mixing it is essential to extract bounds to the flavor conserving couplings.
To do this, we conservatively compared the modification to the couplings of electrons, muons and taus to the $Z$ boson, which have been accurately measured at LEP~\cite{LEP}, with the error in each measurement.
In the triplet case, the effects involve dominantly the left-handed couplings, giving:
\begin{equation}
\left| \frac{\delta g_{Z l^+_L l^-_L}}{g_{Z}^{\rm SM}} \right| = \frac{|v_E^i|^2}{1-2 \sin^2 \theta_W} < \left\{ \begin{array}{l}
0.33 \% \\ 1.23 \% \\ 0.66 \%
\end{array} \right. \quad \mbox{for} \quad \begin{array}{c}
e \\ \mu \\ \tau
\end{array}
\end{equation}
where the listed values correspond to the 3 sigma uncertainties in the measurements (correlations between the measurements in the various flavors are small and have been neglected).
This analysis leads to the following bounds:
\begin{equation}
|v_E^e|^2 < 1.8 \cdot 10^{-3}\,, \quad |v_E^\mu|^2 < 6.8 \cdot 10^{-3}\,, \quad |v_E^\tau|^2 < 3.6 \cdot 10^{-3}\,. \label{eq:bounds2}
\end{equation}
Combining the bounds in Eq.s~(\ref{eq:bounds1}) and (\ref{eq:bounds2}), we can infer that, depending on the flavor structure of the mixing,
\begin{equation}
\lambda < 10^{-3} \div 10^{-5}\,.
\end{equation}
This analysis shows that the mixing needs indeed to be small. This fact has two important consequences on the phenomenology: on one hand, single production processes, proportional to $\lambda$, are very suppressed and can be neglected; on the other hand, the mass splitting between the 3 components in the triplet are very small, therefore decays like $X^0 \to W^+ X^-$ and $X^{-} \to W^+ X^{--}$ are negligible.

 \begin{figure}[h]
\begin{center}
\includegraphics[width=0.42\textwidth]{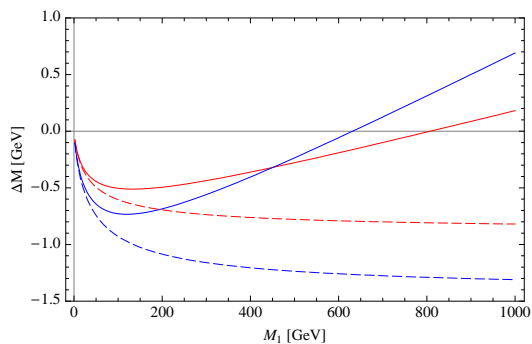}\\
\caption{Mass splitting $M_{X^-} - M_{X^{--}}$ (red) and $M_{X^0} - M_{X^{--}}$ (blue) as a function of the VL mass $M_1$ for maximal $\lambda = 10^{-3}$ (solid lines) and $\lambda = 0$ (dashed).}\label{figure-SpectrumX}
\end{center}
\end{figure}

To be more quantitative, due to the smallness of $\lambda$, the mass splitting can be sensitive to radiative corrections from electroweak gauge bosons, proportional to the breaking of weak isospin.
A simple loop calculation~\cite{Feng:1999fu} leads to (keeping the leading term in an expansion for small $\lambda$):
\begin{eqnarray}
M_{X^-} - M_{X^{--}} &\sim& \lambda M_1 - \frac{\alpha_W M_1}{4 \pi} \left[ (3 \sin^2 \theta_W -1) (f(x_Z)-f(0)) + f(x_W) - f(0) \right]\,, \\
M_{X^0} - M_{X^{--}} &\sim& 2 \lambda M_1 - \frac{\alpha_W M_1}{4 \pi} \left[ 4 \sin^2 \theta_W  (f(x_Z)-f(0)) \right]\,.
\end{eqnarray}
In the above formulas $\alpha_W = \alpha/\sin^2 \theta_W$, $x_W = m_W/M_1$, $x_Z = m_Z/M_1$, the loop function is given by $f (x) = 2 \int_0^1 dy\; (1+y) \log (y^2 + (1-y) x^2)$, and $f(0)$ is the contribution of the photon loop.
As a check, for unbroken weak isospin, i.e. $m_W = m_Z = m_\gamma$, the loop induced splitting vanishes; furthermore, for large $M_1$ we have that $f(x)-f(0) \sim 2 \pi x$ thus the loop correction is proportional to the $W$ and $Z$ masses and not $M_1$. Another interesting point is that the sign of the loop corrections is opposite to the contribution of $\lambda$, and for any value of $\lambda$ its contribution will become sub-dominant for large enough $M_1$.
To quantify this statement, in Fig.~\ref{figure-SpectrumX} we show the mass splitting as a function of $M_1$ for the maximum allowed value of $\lambda = 10^{-3}$ (solid lines) and for $\lambda = 0$ (dashed lines).
We can see that the ordering of masses changes for largish $\lambda$, and we checked that for $\lambda < 0.5\cdot 10^{-4}$ the electroweak loops always dominate for masses below a TeV.
We can also see that for any allowed value of $\lambda$ the mass splittings are small and of the order of 1 GeV, thus they can be safely neglected in the LHC phenomenology studies.
Furthermore, it has been observed in~\cite{Delgado:2011iz} that for $\lambda > 10^{-12}$ the two body decays always dominate, and in the following we will focus on this scenario.

%%%%%%%%%%%%%%%%%%%%%%%%%%%%%%%%%%%
\section{LHC Phenomenology of the Lepton Triplet}

In order to study the phenomenology for the exotic leptons, we firstly consider their decays in the triplet model. The decay widths and branching fractions depend on the masses and the couplings of the exotic
leptons. For these degenerate massive exotic leptons whose masses are above the scale of $M_W$, the dominant decay modes are $X\rightarrow \ell + W, Z \,{\rm or}\, H$. The partial decay widths are proportional to the square of the mixing parameters $v_i$, so that by summing over the flavors of the lepton in the final state, we obtain the parameter $\lambda$.
The total decay widths of the exotic triplet leptons are

\begin{eqnarray}
&&\Gamma_{X^{--}} = \frac{G_{F}M^3}{4\sqrt{2}\pi} \lambda \; F_{1}(r_{W})\,, \nonumber\\
&&\Gamma_{X^{-}} = \frac{G_{F}M^3}{16\sqrt{2}\pi}\lambda \; (2F_{1}(r_{W})+F_{1}(r_{Z})+F_{0}(r_{H}))\,, \\
&&\Gamma_{X^{0}} = \frac{G_{F}M^3}{8\sqrt{2}\pi}\lambda \; (F_{1}(r_{Z})+F_{0}(r_{H}))\,. \nonumber
\end{eqnarray}
Here, the function $F_n$ are defined as
\begin{equation}
 F_{n}(r)=(1+2nr)(1-r)^2,
 \end{equation}
where $r_{a}=m_{a}^2/M_{X}^2$.
It is clear thus that $F_n (r_W)$ is generated by the partial width into $W + l$, and so on, and that the rates into the various bosons $W$, $Z$ and $H$ are only determined by the representation the heavy leptons belong to.
On the other hand, the rates in the various lepton families are determined by ratios of $|v_i/v_j|^2$, and all branching ratios are independent on $\lambda$.

\begin{figure}[h]
\begin{center}
\includegraphics[width=0.5\textwidth]{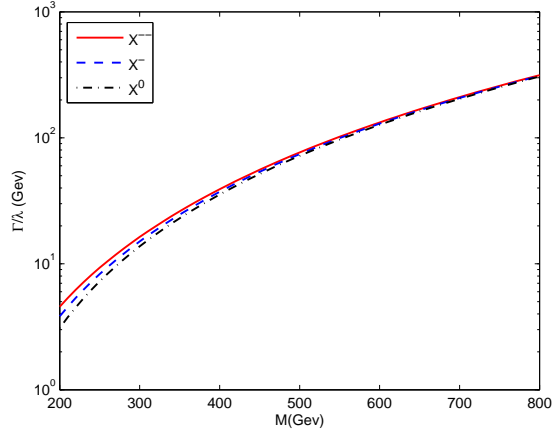}\\
\caption{Decay widths of the exotic triplet leptons as a function of the mass and in units of $\lambda$.}\label{figure-TripletXdecaywidth}
\end{center}
\end{figure}

The squared sum of the mixing angles $\lambda$ is of fundamental importance for the decay life times, and to determine if 3-body decays are negligible: we are interested in the case $\lambda> 10^{-12}$, where the exotic lepton leading decay modes are decays to the standard model leptons, and decays between two heavy leptons, like $X^{--} \to W^* X^-$ are negligible.
The formulas for the leading decay channels are listed in Table~\ref{table100-1}, while the total decay widths of the triplet leptons in units of $\lambda$ are plotted in Fig.~\ref{figure-TripletXdecaywidth}.
In the parameter range of interest, the decays are always prompt and we do not expect displaced vertices nor long-lived charged tracks.

 \begin{table}[h]
\begin{center}
\begin{tabular}{|c|c|}
\hline
channel & Partial Width /$\frac{G_{F}M^3}{16\sqrt{2}\pi}$ \\
\hline
$X^{--} \to e_{i}^{-} W^{-}$ & $4F_{1}(r_{W})\mid v_{Ei} \mid^2$ \\
\hline
$X^{-} \to e^{-}_{i}H $ & $F_{0}(r_{H})\mid v_{Ei} \mid^2$ \\
$X^{-} \to e^{-}_{i}Z $ & $F_{1}(r_{Z})\mid v_{Ei} \mid^2$ \\
$X^{-} \to \nu_{i}W^{-} $ & $F_{1}(r_{W})\mid v_{Ni} \mid^2$ \\
\hline
$X^{0} \to \nu_{i}H $ & $F_{0}(r_{H})\mid v_{Ni} \mid^2$ \\
$X^{0} \to \nu_{i}Z $ & $F_{1}(r_{Z})\mid v_{Ni} \mid^2$ \\
\hline
\end{tabular}
\end{center}
\caption{Allowed decay channels and partial widths for the triplet exotic leptons.}\label{table100-1}
\end{table}

\begin{figure}[h]
\begin{center}
\includegraphics[width=0.42\textwidth]{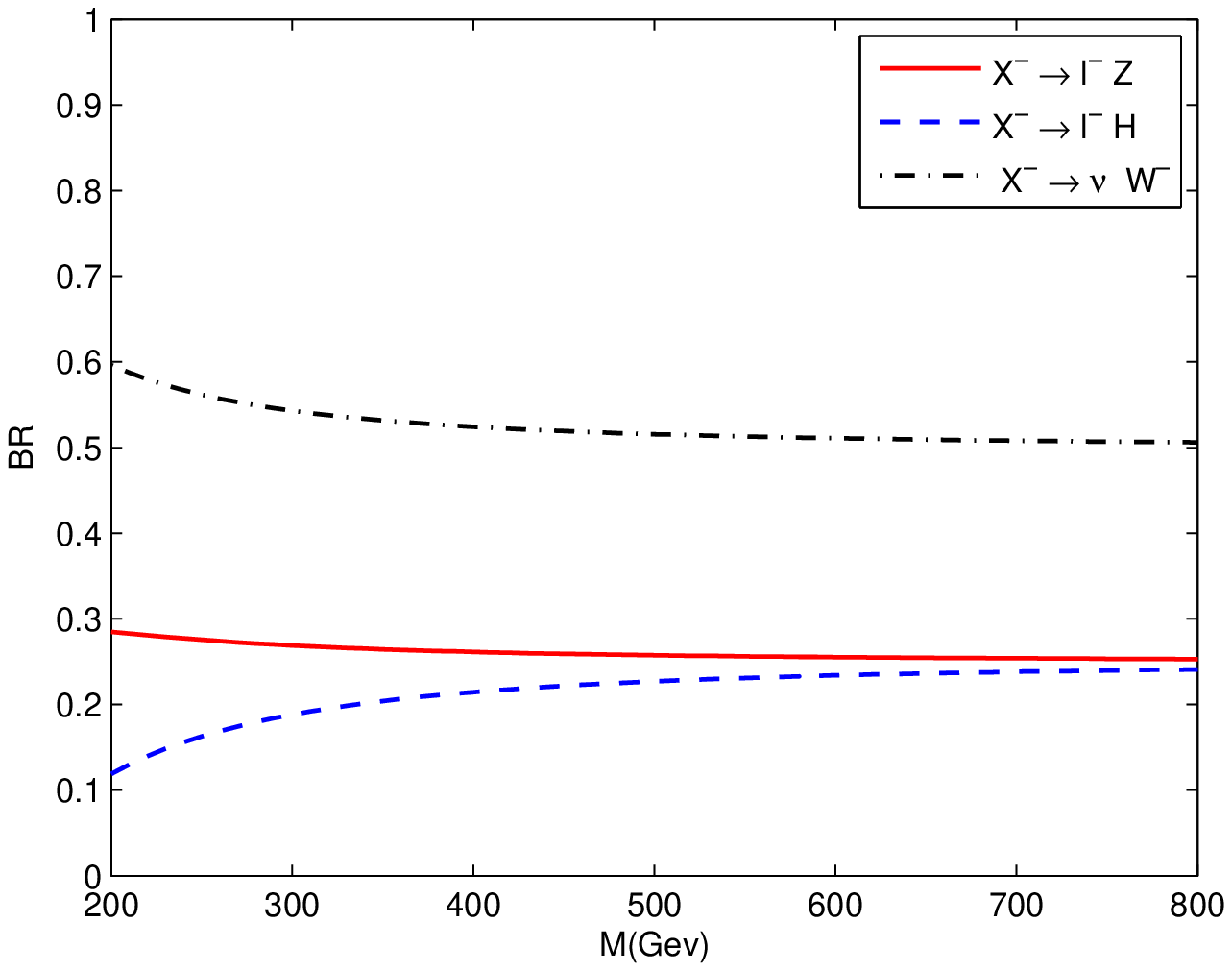}\includegraphics[width=0.45\textwidth]{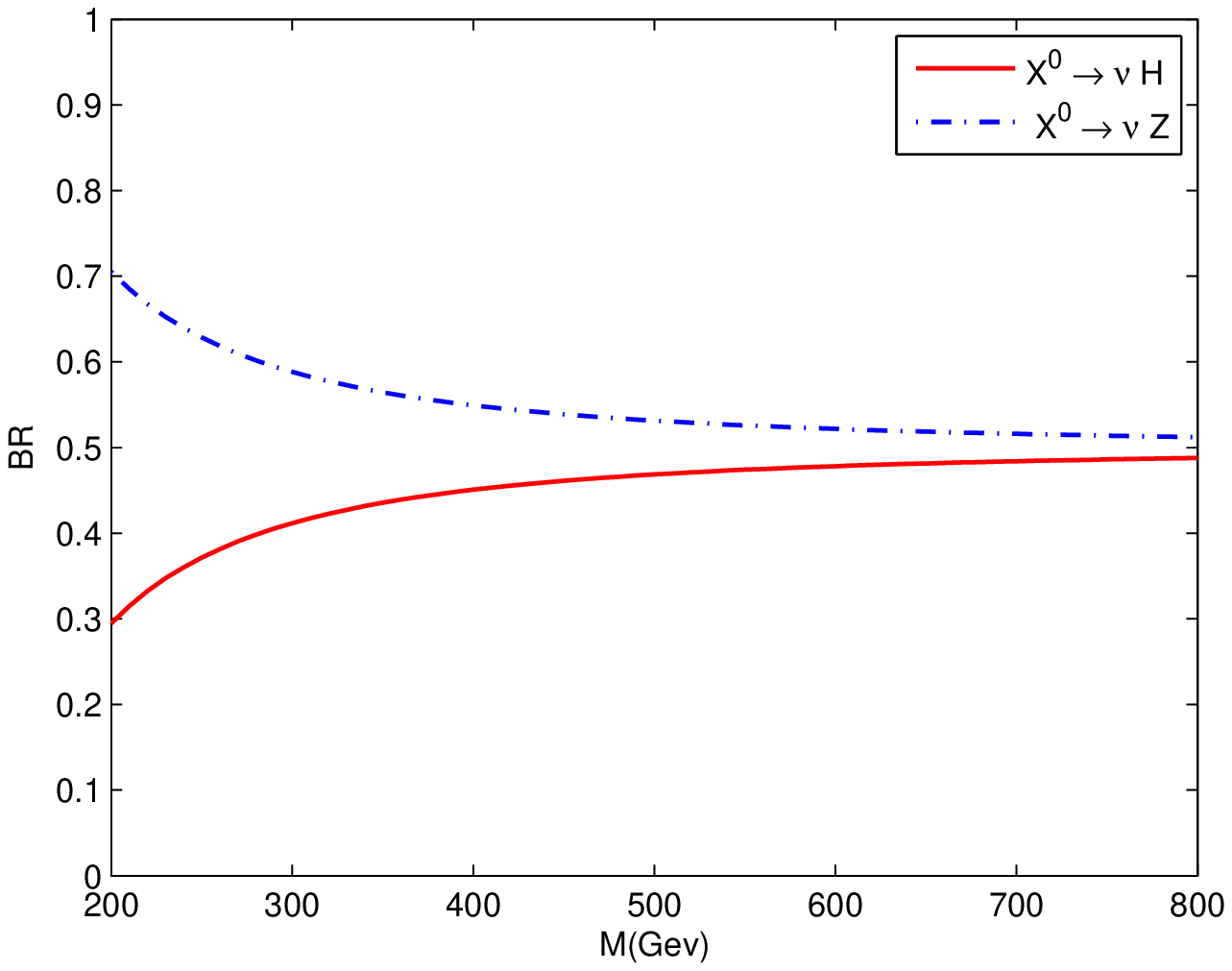}\\
\caption{The decay branching fracions of $X^{-}$ and $X^{0}$ in exotic lepton triplet }\label{figure-TripletXdecaybranch}
\end{center}
\end{figure}

The doubly charged lepton $X^{--}$ can only decay to a charged lepton $\ell^-$ and a $W^-$ with a $100\%$ branching fraction; $X^-$ and $X^0$ have the branching fractions to $W$, $Z$ and $H$.
We show the branching fractions of $X^-$ and $X^0$ in  Fig.~\ref{figure-TripletXdecaybranch}.

If the mixing parameter $\lambda$ is not very small ($\lambda> 10^{-12}$), the determination of the leptonic decay mode would bring the interesting phenomenology. The relative fraction of the exotic leptons decay to a specific charged lepton depends on the ratio of the mixing angles squared $ |v_e|^2$, $ |v_\mu|^2$ and $ |v_\tau|^2$. In order to simplify the discussion, we followed Ref.~\cite{Delgado:2011iz} and assumed that $|v_e|^2=|v_\mu|^2=|v_\tau|^2$ as a basic example. Any different relative branching fraction determinted by specific mixing models could be parameterized as a factor on the signal product rates, and our analysis extended by a simple rescaling of the signal rates.
For later reference, we can define a parameter $\zeta_\tau = |v_\tau|^2/\lambda$ which encodes the decay rates into taus, that we fix to $\zeta_\tau = 1/3$ in our simulations.

\begin{figure}[h]
\begin{center}
\includegraphics[width=0.6\textwidth]{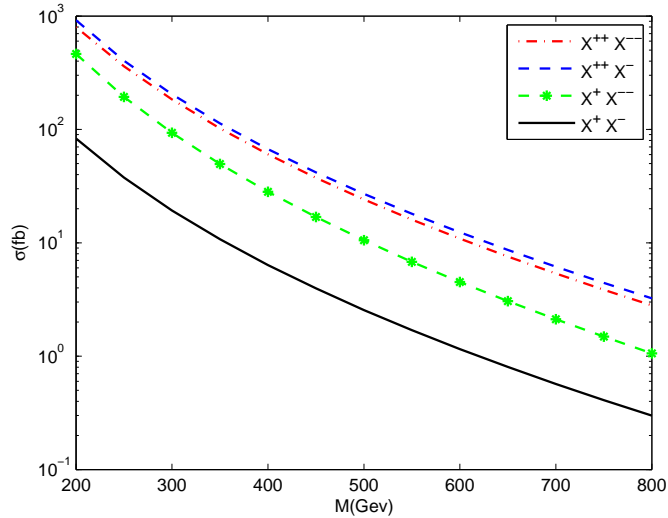}\\
\caption{Total cross sections of triplet lepton pair or associated productions at the 14 TeV LHC}\label{figure-prodrate21}
\end{center}
\end{figure}

The exotic leptons in the triplet can be pair produced via their interactions to the electroweak gauge bosons, which are fixed by the representation of the exotic leptons.
In the following we will be mainly interested in the phenomenology of the doubly and singly charged leptons, because they always contain a charged lepton in the final state.
The total cross sections at 14 TeV in various channels as a function of the lepton masses are plotted in Fig.~\ref{figure-prodrate21}.
We can see that these productions of the exotic leptons have sizable rates, because they are produced via the standard model electro-weak gauge interaction and they are not suppressed by mixing angels.
Single production in association with a SM lepton, being proportional to $\lambda$, is always negligibly small~\cite{Delgado:2011iz}.

In the following we will discuss the most promising channels for the discovery of the exotic leptons, focusing on a 14 TeV LHC run. We chose this center of mass energy mainly to be able to compare our results with the ones in~\cite{Delgado:2011iz}, knowing that the next LHC run will likely be at a lower energy of 12.5 or 13 TeV.
While the rates will be reduced for both signal and background, our main conclusions will be unchanged.
Furthermore, generic searches for lepton-rich final states at 8 TeV can already pose significant bounds on the masses of the exotic leptons: for instance both ATLAS~\cite{ATLAS} and CMS~\cite{CMS} have preliminary results for searches of 3 and 4 lepton final states, with bounds on the fiducial cross section in the typical range $1\div 10$ fb.
Considering the rate loss due to the $W$ and $Z$ leptonic decays, we can expect bounds on the exotic lepton masses below $300$ GeV (for low masses the cross sections at 8 TeV are about 1/3 of the ones at 14 TeV). A much higher reach can indeed be obtained by dedicated searches, as we will show in the following.

\subsection{Doubly charged lepton detection at LHC via $X^{++}X^{--}$ pair production and $X^{++}X^{-}/X^{+}X^{--}$ associated production}

The doubly charged leptons are the most characteristic particles in the model.
They can be pair produced $X^{++}X^{--}$ via an s-channel $Z$ and photon, and in $X^{++}X^{-}/X^{+}X^{--}$ associated production via an s-channel $W$.
They dominantly decay into a SM charged lepton with a same-sign $W$ boson, therefore leading to a pair of same-sign leptons via the $W$ boson leptonic decay, giving a characteristic track of the doubly-charged nature of the new particle. This channel has been carefully studied in Ref.~\cite{Delgado:2011iz}, so we follow the same method to analyze the $X^{++}X^{--}$ pair production channel. We however extended the previous study by adding the contribution of the irreducible associated production channel $X^\pm X^{\mp \mp}$, which will significantly increase the signal rate.

We will consider only $e$ and $\mu$ in our definition of a lepton for the sake of experimental identification, leaving tau final states for a later work. Furthermore, one of the $W$ bosons in the pair production final state is required to decay leptonically for the charge identification and the other W to
decay hadronically for the mass reconstruction. As a result, the final state for the doubly charged lepton $X^{++}X^{--}$ pair production channel is

\begin{equation}
pp\rightarrow X^{--}X^{++}\rightarrow \ell^- W^-\ell^+W^+ \rightarrow \ell^- \ell^+ jj \ell^-\bar{\nu}(\ell^+\nu)\nonumber
\end{equation}
with an inclusive branching fraction
\begin{equation}
BR=(1-\zeta_\tau)^2\times(0.68)\times(0.21)\times 2 \approx 0.13
\end{equation}
Here, we simply let the branching fractions of $X^{--}$ decay to leptons be the same, $\frac{1}{3}$, as a typical case.
As one of the final W bosons decays to hadrons in the $X^{++}X^{--}$ pair production signature, and the measurement accuracy of the hadronic calorimeter is not enough to distinguish the $W$ or $Z$ bosons, the following process will provide additional signature at LHC with the same final states:

\begin{equation}
pp\rightarrow X^{--}X^{+} +h.c.\rightarrow \ell^- W^-\ell^+Z + h.c. \rightarrow \ell^- \ell^+ jj \ell^-\bar{\nu}(\ell^+\nu)
\end{equation}

The mass of singly changed lepton $X^- $ is almost the same as doubly changed lepton $X^{--}$, therefore the two resonances overlap when we reconstruct their invariant masses. Thus we need to consider the two channels together.
The branching ratios in this production channel is
\begin{equation}
BR=(1-\zeta_\tau)^2\times(0.70)\times(0.21)\times \mbox{BR}_{X^-\to Z} \approx 0.06 \times \mbox{BR}_{X^-\to Z}
\end{equation}
therefore it will give a rate of events which is about $1/8$ of the pair production case as $ \mbox{BR}_{X^-\to Z} \sim 0.25$.

In this channel, we define the signal identification cuts in the same way as in Ref.~\cite{Delgado:2011iz}: three charged lepton and two jets, plus missing energy. Considering the
detector coverage for the LHC experiments, we apply the following basic kinematical acceptance cuts on the
transverse momentum $p_T$, pseudo-rapidity $y$, missing transverse energy $E\!\!\!\slash_T$, and the particle separation $\Delta R$:
\begin{eqnarray}
&&p_T(\ell) > 15 {\,\rm GeV},\,|\eta(\ell) | < 2.5,\, E\!\!\!\slash_T>25 {\,\rm GeV},\nonumber\\
&&p_T(j) > 15 {\,\rm GeV},\,|\eta(j) | < 2.5,,\nonumber\\
&&\Delta R(ll) > 0.3,\,\Delta R(jl) > 0.4,\,\Delta R(jj) > 0.4. \label{eq_basiccuts}
\end{eqnarray}
The jets in the signal come from $W$ or $Z$ boson decays, so we require their invariant mass to be in the $W/Z$ mass window:
\begin{equation}
 M_W -20\,\rm GeV < M(jj)  < M_Z + 20 \,\rm GeV.
\end{equation}

To significantly reduce the backgrounds, it is essential to reconstruct the mass of the heavy leptons and, as the doubly-charged one decays into two same sign leptons, one needs to rely to the correct charge identification of the leptons in the final states, which is very effective for electrons and muons of moderate momentum.
One resonance can be easily identified by associating the jets with the lepton of different charge, thus $M(\ell^\pm jj) \sim M_1$. There is missing energy in the remaining two same sign charged leptons and one neutrino, so reconstructing the other exotic lepton mass is not straightforward. However, as there is a single massless missing particle in the event, we can solve for all the possible neutrino momenta that would come from a $W$ boson decay and choose the solution which gives the closest $M(\ell^\mp\ell^\mp \bar{\nu})$ to $M(\ell^\pm jj)$. The scheme efficiency of the $M(\ell^\mp\ell^\mp \bar{\nu})$ reconstruction is $97\%$~\cite{Delgado:2011iz}. Therefore, we can reconstruct the two exotic leptons with same mass. To suppress the background and strengthen the signal observation, it is useful to devise the following mass cut
\begin{equation}
| M(jj\ell^\pm) - M( \ell^\mp \ell^\mp \bar{\nu}) | < 30 \,\rm GeV.
\end{equation}

There are still some standard model backgrounds that lead to the same final state as our signal. We used MadGraph5~\cite{MadGraph} to calculate the irreducible backgrounds $ l^+ l^- j j W^{\pm} $ and $ W^+ W^- j j W^{\pm}  $ where  the W bosons decay leptonically.
We do not include other reducible backgrounds, like $Z + jets$ where one jet is misindentified as a lepton and $t \bar{t}$ with a third lepton coming from the $b$ decays, which are found to be important~\cite{ATLAS,CMS} in multi-lepton searches.
Their rates can only be estimated from the data and are therefore beyond the scope of our work, furthermore their impact is found to be smaller in the regions with high $p_T$ leptons which is where our signal should cluster.
We list the cross sections for the signal and irreducible backgrounds after the cuts in Table~\ref{table-effect}. We can see that the $W$ mass cut, and especially the $X$ invariant mass matching, are very effective in reducing the backgrounds while preserving the signal events.

\begin{table}[h]
\begin{center}
\begin{tabular}{|c|c|c|c|}
\hline
process & basic cuts & $W$ mass cut & $X$ mass matching \\
\hline
$X^{--}X^{++}$ & 28 & 28 & 27 \\
$X^{++}X^{-}$ & 6.4 & 6.4 & 6.2 \\
$X^{--}X^{+}$ & 3.6 & 3.6 & 3.5 \\
\hline
$ l^+ l^- j j W^{\pm}  $ &$ 78$&$23.5$& $ 3.6 $\\
$ W^+ W^- j j W^{\pm}  $ & $ 0.24 $&$0.036$&$0.036$\\
\hline
\end{tabular}
\end{center}
  \caption{Effect of the kinematical cuts on the production cross section (fb) at the 14 TeV LHC for the signal ($M_X=200$ GeV) and backgrounds.}\label{table-effect}
\end{table}

\begin{figure}
\begin{center}
\includegraphics[width=0.45\textwidth]{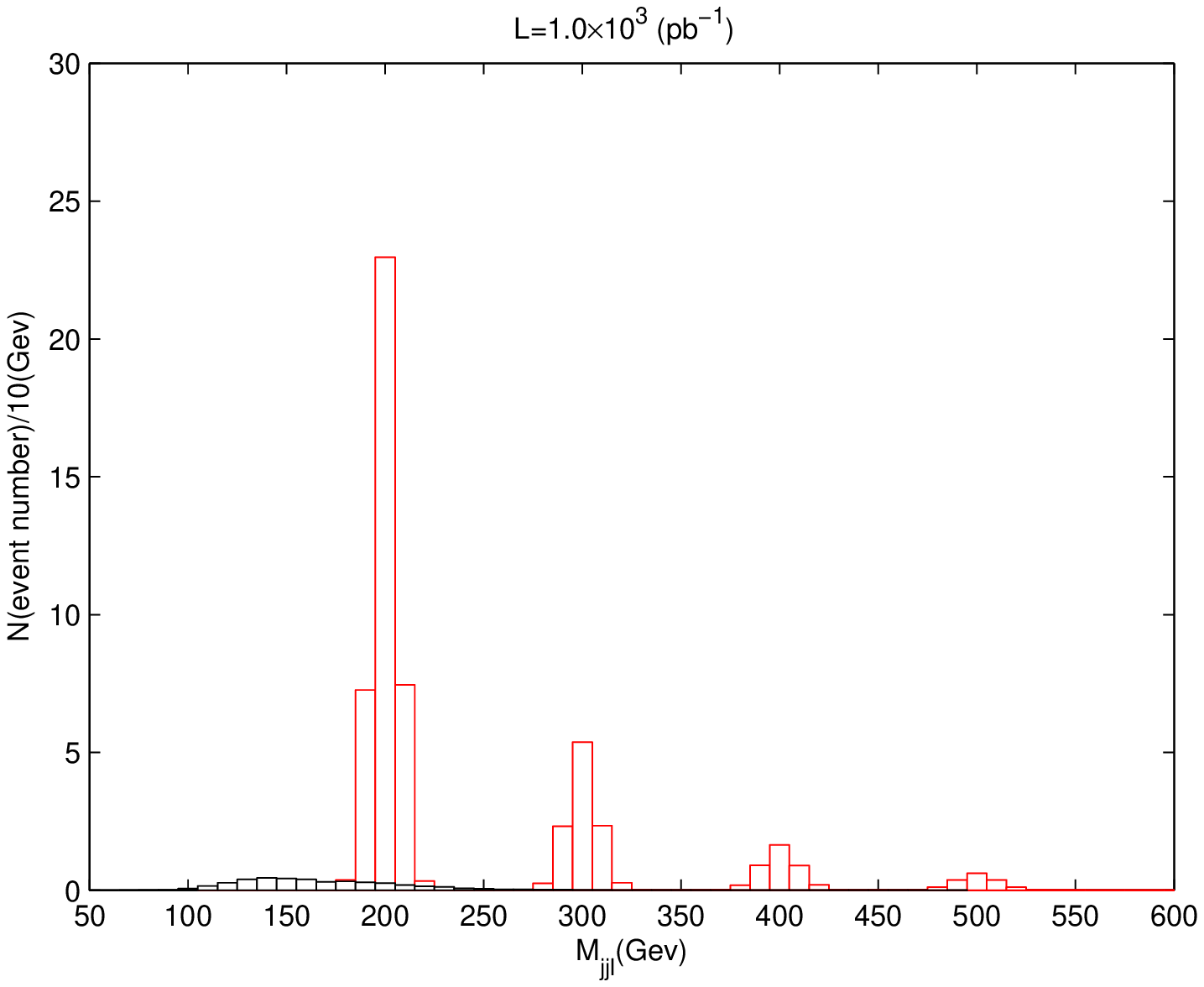}
\includegraphics[width=0.45\textwidth]{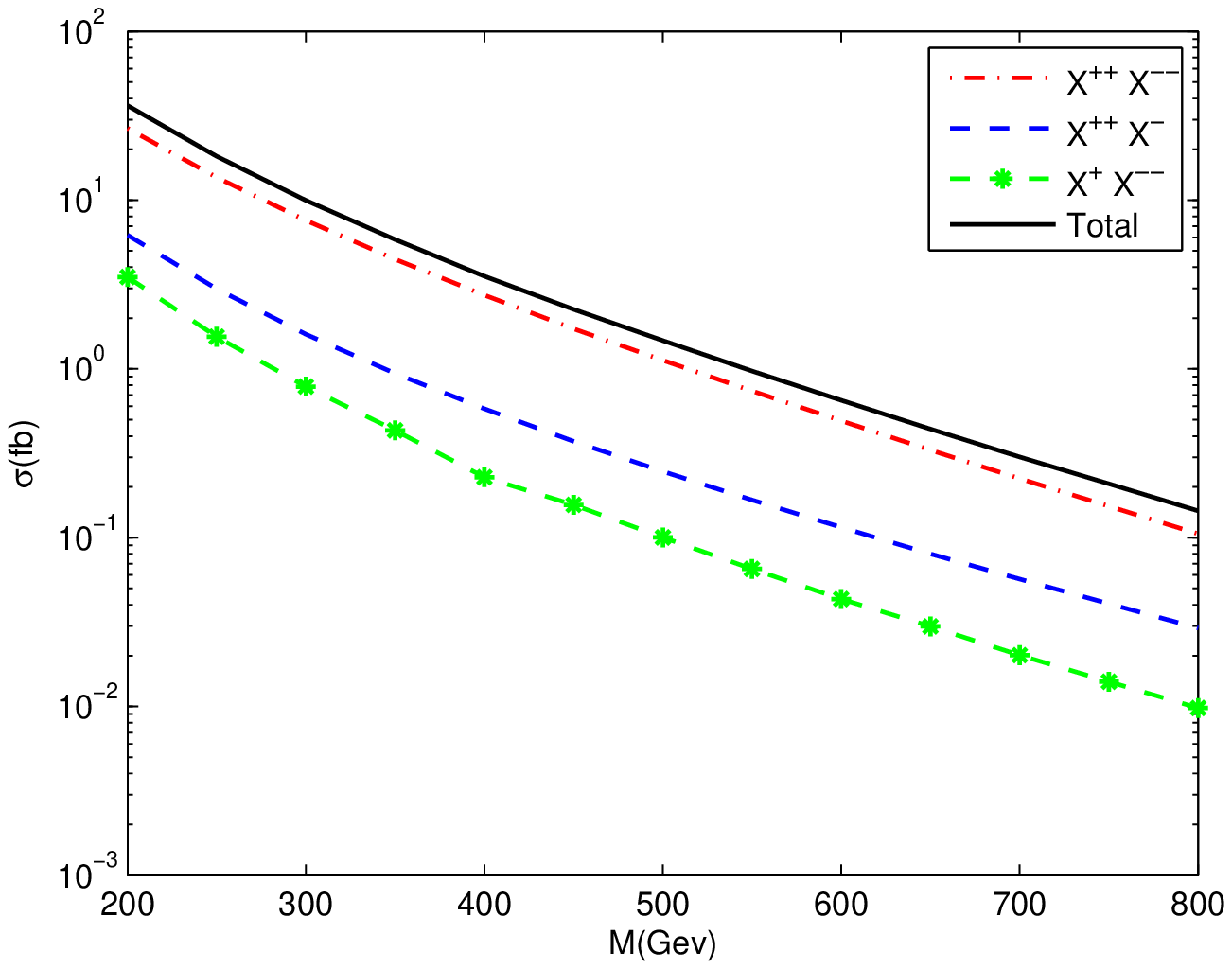}\\
\caption{Left: invariant mass $M(\ell jj)$ distribution curves of exotic leptons signal resonances and the background. Right: signal cross section in the $\ell^- \ell^+ jj \ell^-\bar{\nu}(\ell^+\nu)$ channel at 14 TeV LHC, after the cuts.}\label{figure-ljjdis}
\end{center}
\end{figure}

In Fig.~\ref{figure-ljjdis}, we plotted the distribution of the invariant mass $M(\ell jj)$ for exotic lepton signals at different masses and the background. The resonance widths in the Fig.~\ref{figure-ljjdis} are much broader than the physical decay widths, because we smeared the lepton and jet energies with a Gaussian distribution  as in Ref\cite{Delgado:2011iz} to simulate the detector sensitivity according to
\begin{equation}
\frac{\delta E}{E}=\frac{a} {\sqrt{E/ \mbox{GeV}}}\oplus b,
\end{equation}
\noindent
where $a = 5\%$, $b = 0.55\%$ for leptons and $a = 100\%$, $b = 5\%$ for jets \cite{Ball:2007zza}. The width is widened with the exotic lepton mass, and we only take into account the events within
  \begin{equation}
	 |M(\ell jj)-M_X|<0.1 M_X.
\end{equation}
The surviving SM background events are distributed at the low invariant mass region where the resonance width is narrow, thus the background is negligible if we only consider the events within the signal resonances.

After considering the basic acceptance cuts and the small semi-leptonic decay branching fractions of the exotic leptons, the surviving signal cross section is still sizable when the mass of lepton triplet is not larger than 500 GeV as showed in the right panel of Fig.\ref{figure-ljjdis}. Compared to only considering the $X^{++}X^{--}$ channel, the additional $X^{++}X^{-}/X^{+}X^{--}$ channel can contribute considerable signal cross section and increase it by about one third.

\begin{figure}
\begin{center}
\includegraphics[width=0.6\textwidth]{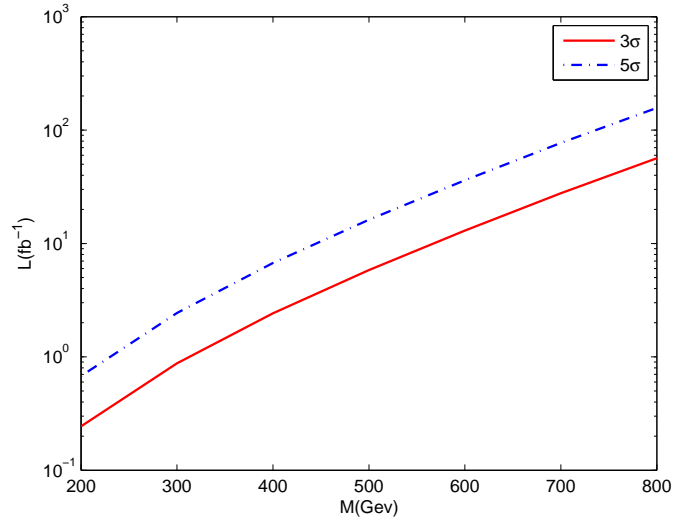}\\
\caption{The needed luminosity to observe different mass triplet leptons via $\ell^- \ell^+ jj \ell^-\bar{\nu}(\ell^+\nu)$ channel for 3 $\sigma$ and 5 $\sigma$ significances at the 14 TeV LHC}\label{figure5-2}
\end{center}
\end{figure}

In order to quantify the LHC reach, we define the statistical significance $s$ as
\begin{equation}
s=\frac{N_{s}}{\sqrt{N_{s}+N_{b}}} ,
\end{equation}
where $N_s$ is the number of signal events, and $N_b$ is the number of background events. If the signal cross section $\sigma_s$ and background  cross section $\sigma_b$ are obtained, we can derive the needed integrated luminosity $L$ as a function of the significance
\begin{equation}
 L = s^2 \left( \frac{\sigma_{s}+\sigma_{b}}{\sigma_s^2} \right).
\end{equation}
We then can calculate the needed integrated luminosity at LHC to reach a certain statistical significance.
The needed integrated luminosity  to observe different mass triplet leptons is plotted in Fig.~\ref{figure5-2}.
Because we have already considered the additional signal events contributed by $X^{--}X^+$ and $X^{++}X^-$ associated productions, the luminosity needed to discover lepton triplet in this channel is reduced compared to the result in~\cite{Delgado:2011iz}.

\subsection{Singly charged lepton detection via $X^{++}X^{-}/X^{+}X^{--}$ associated production and $X^{+}X^{-}$ pair production}

Although the associated production can contribute the same signal as the $X^{--}$ pair production to detect the exotic doubly charged leptons, this channel also provides special final states which can signal the presence of a singly-charged exotic lepton. The decay modes of the singly charged lepton $X^{\pm}$ are more complex than the doubly charged one. While the decay mediated by the $W$ has a neutrino, thus less visible particles, the decays via $Z$ and $H$ offer new potentially interesting channels. We first consider the decay to $Z$ boson and SM leptons, whose branching ratio is about $25\%$, see Fig.~\ref{figure-TripletXdecaybranch}. Then we let the $Z$ boson decay to lepton pairs and the $W$ bosons decay to hadrons. So there are four leptons and two jets in the signal final states, as
\begin{equation}
pp\rightarrow X^{--}X^{+} +h.c.\rightarrow \ell^- W^-\ell^+Z + h.c. \rightarrow \ell^- \ell^-\ell^+ \ell^+ jj
\end{equation}

In order to obtain large cross section, we required the $W$ boson from the doubly charged lepton to decay in hadrons  in the $X^{++}X^{-}/X^{+}X^{--}$ associated production. As the measurement accuracy of the hadronic calorimeter is not enough to distinguish the $W$ or $Z$ bosons, the following  $X^{+}X^{-}$ pair production process will provide additional signature at LHC with the same final state:

\begin{equation}
pp\rightarrow X^{+}X^{-} +h.c.\rightarrow \ell^- Z \ell^+Z + h.c. \rightarrow \ell^- \ell^-\ell^+ \ell^+ jj
\end{equation}

The mass of the singly changed lepton $X^-$ is almost the same as doubly changed lepton $X^{--}$, so we should consider the two channel together even though the $X^-$ pair production is small compared to the associated production, see Fig.~\ref{figure-prodrate21}.

We also apply the same basic kinematical acceptance cuts on the transverse momentum, rapidity, and particle separation as in Eq.~(\ref{eq_basiccuts}). As there are no neutrinos in the signal events, we replace the missing energy cut with a veto cut $ E\!\!\!\slash_T < 25 {\,\rm GeV}$.
The $Z$ boson in the final state can be tagged by lepton pair reconstruction, therefore we require the invariant mass of a lepton pair to be close to the $Z$ boson mass, within:
\begin{equation}
| M(\ell^+\ell^-) - M_Z | < 5 \,\rm GeV.
\end{equation}
Because there are four leptons (possible two lepton pairs) in the final states, we should consider all the possible combinations. If one of the combinations has the lepton pair in the $Z$ boson width, the event will pass the cut, and we treat the lepton pair as $Z$ boson.

In the following, therefore, we only consider backgrounds which include a leptonic $Z$ boson.
The backgrounds are
\begin{eqnarray}
pp\rightarrow \ell^+\ell^-  jj Z(\ell^+\ell^-)\,,\qquad
pp\rightarrow W^+(\ell^+\nu)  W^- (\ell^-\bar{\nu})    jj Z(\ell^+\ell^-)\,.
\end{eqnarray}
As before, we also require the $W$ mass cut on the jets
\begin{equation}
| M(jj) - M_W | < 20 \,\rm GeV.
\end{equation}

All the particles in final state are visible, so it is straightforward to reconstruct the two exotic leptons and require their masses are similar in the following range:
\begin{equation}
| M(jj\ell^-) - M( \ell^+ Z(\ell^+ \ell^-))  |\; \mbox{or}\; | M(jj\ell^+) - M( \ell^- Z(\ell^+ \ell^-))  |  < 20 \,\rm GeV.
\end{equation}

\begin{table}[h]
\begin{center}
\begin{tabular}{|c|c|c|c|}
\hline
process & basic cuts & $W$ mass cut & $X$ mass matching \\
\hline
$X^{++}X^{-}$ & 1.9 & 1.9 & 1.9 \\
$X^{--}X^{+}$ & 1.1 & 1.1 & 1.1 \\
$X^{+}X^{-}$ & 0.1 & 0.1 & 0.1 \\
\hline
$ \ell^+ \ell^- j j Z  $ &$ 25.5$&$2.6$& $ 1.9 $\\
$ W^+ W^- j j Z  $ & $ 0.35 $&$0.02$&$0.0$\\
\hline
\end{tabular}
\end{center}
  \caption{Effect of the kinematical cuts on the production cross section (fb) at the 14 TeV LHC for the $\ell^+\ell^-  jj Z(\ell^+\ell^- )$ signal ($M_X=200$ GeV) and backgrounds.}\label{table-effect2}
\end{table}

\begin{figure}
\begin{center}
\includegraphics[width=0.45\textwidth]{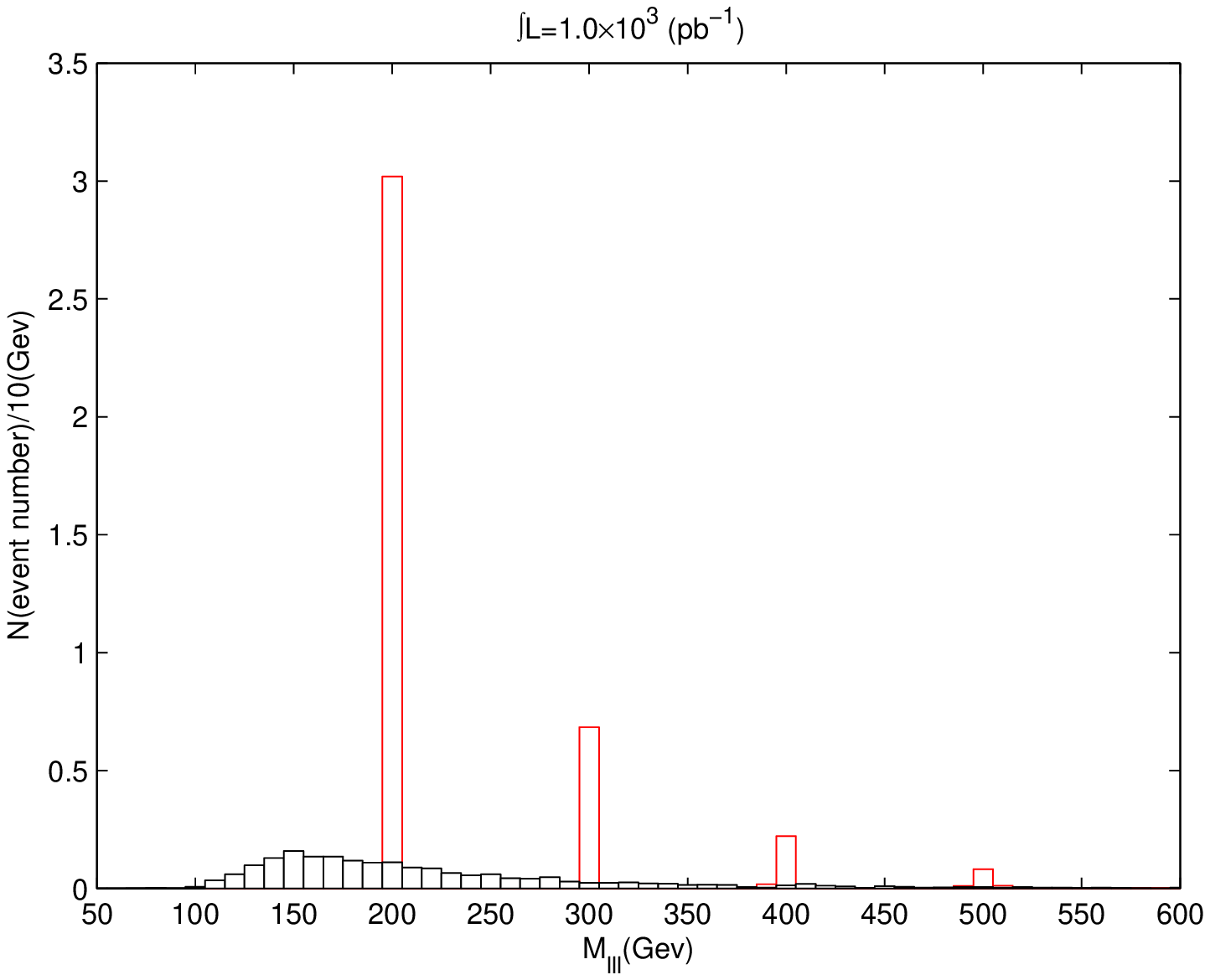}
\includegraphics[width=0.45\textwidth]{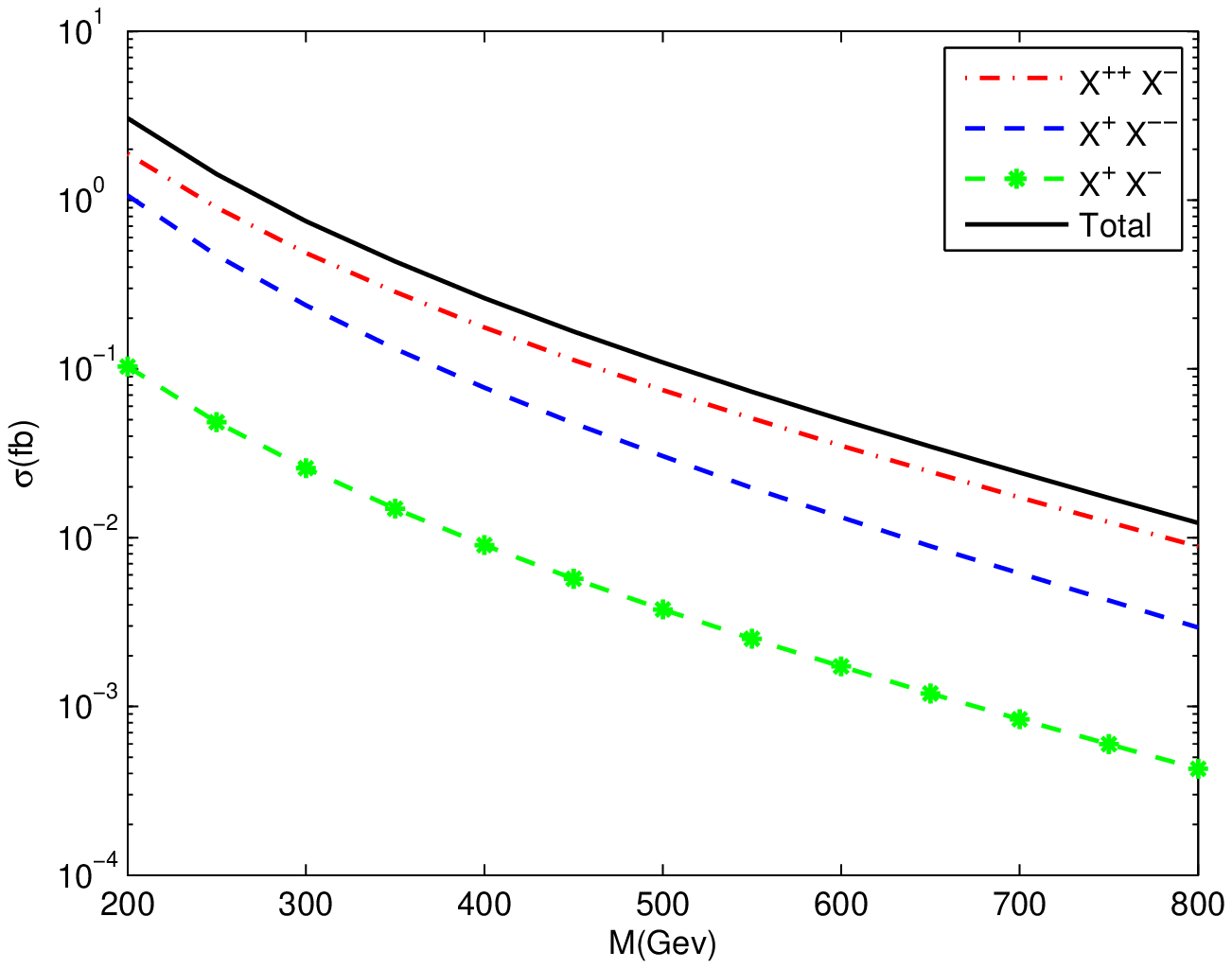}\\
\caption{Left: invariant mass $M(\ell \ell\ell)$ distribution for exotic lepton signal at various masses and the background. Right: final lepton triplet signal production rate for the $\ell^+\ell^-  jj Z(\ell^+\ell^- )$ channel at 14 TeV LHC.}\label{figure-llldis}
\end{center}
\end{figure}

The cross sections for the signal and backgrounds, after cuts, are listed in Table~\ref{table-effect2}.  We can see that the two mass cuts are very effective to retain signal and suppress the SM backgrounds. The additional contribution from $X^{+}X^{-}$ pair production process is very small, because the $X^{+}X^{-}$ pair production rate is small compared to the $X^{++}X^{-}/X^{+}X^{--}$ associated production rates as we can see from Fig.~\ref{figure-prodrate21}.
Although the remaining background is still sizeable compared to the signal, it can not reconstruct a resonance.
We can clearly see this in Fig.\ref{figure-llldis}, where we plotted the invariant mass $M(\ell \ell\ell)$ distribution for signals at various masses and the background. The resonances in the Figure are very narrow, and we also see that the SM background events are distributed in the low invariant mass region.
The background contamination can therefore be reduced by limiting the signal region to large invariant masses.
After considering the basic cuts and the small decay branching fractions of the exotic leptons, the final signal cross section is still sizable when the mass of the lepton triplet is not larger than 300 GeV as showed in the right panel of Fig.\ref{figure-llldis}.

\begin{figure}
\begin{center}
\includegraphics[width=0.6\textwidth]{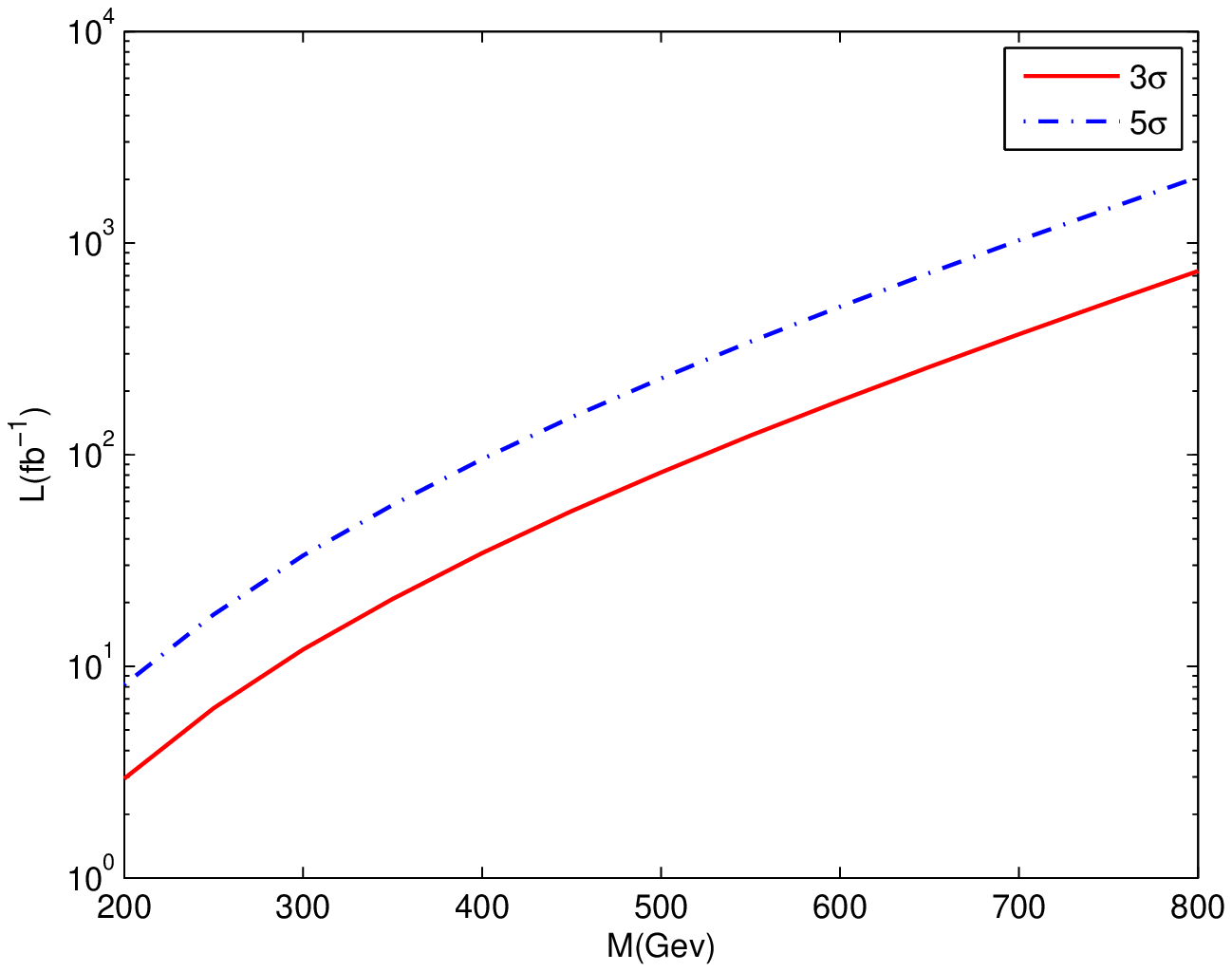}\\
\caption{The needed luminosity to observe different mass triplet leptons via $\ell^+\ell^-  jj Z(\ell^+\ell^- )$ channel for 3 $\sigma$ and 5 $\sigma$ significances at the 14 TeV LHC}\label{figure-lum2}
\end{center}
\end{figure}

The integrated luminosity needed to observe different mass triplet leptons is plotted in Fig.~\ref{figure-lum2}: we can see that this channel is less sensitive than the previous one, as shown in Fig.~\ref{figure5-2}, however it should be stressed that this channel is more sensitive to the production of the singly-charged exotic lepton.

In the $X^{++}X^{-}/X^{+}X^{--}$ production process, the $Z$ bosons from the exotic singly-charged leptons can also decay to $b\bar{b}$ pairs. The branching ratio is larger than the $\ell^+\ell^-$ mode, however we should consider the b-tagging efficiency and more complex backgrounds. The final state contains two charged leptons, two light jets and two b-jets, $\ell^+\ell^-  jjb\bar{b}$, where the two light jets come from a $W$ decay and the two b-jets come from $Z$ decays. We therefore apply the following mass cuts:

\begin{eqnarray}
| M(jj) - M_W | < 20 \,{\rm GeV},\\
| M(b \bar{b}) - M_Z | < 20 \,\rm GeV.
\end{eqnarray}

We also have the mass matching cut for exotic leptons in this channel
\begin{equation}
| M(jj\ell^-) - M( \ell^+ Z(b \bar{b}))  |\; \mbox{or}\; | M(jj\ell^+) - M( \ell^- Z(b \bar{b}))  |< 20 \,\rm GeV.
\end{equation}

We considered the SM backgrounds $\ell^+\ell^-  jjb\bar{b}$ and $W^+W^-  jjb\bar{b}$, and we found that the latter is negligible. However the $\ell^+\ell^-  jjb\bar{b}$ background is not small even after all the mass cuts, as shown in Table~\ref{table-effect3}. We find the $\ell^+\ell^-$ pairs in background are dominantly from on-shell $Z$ boson decay, so we impose a $Z$ mass veto cut:
\begin{equation}
| M(\ell^+\ell^-) - M_Z | > 5 \,\rm GeV.
\end{equation}
to suppress the background.
\begin{table}[h]
\begin{center}
\begin{tabular}{|c|c|c|c|c|}
\hline
process & basic cuts & $W/Z$ mass cut & $Z$ mass veto &$X$ mass matching \\
\hline
$X^{++}X^{-}$ & 2.1 & 2.1 & 1.9 & 1.9 \\
$X^{--}X^{+}$ & 1.1 & 1.1 & 1.0 & 1.0 \\
$X^{+}X^{-}$ & 0.1 & 0.1 & 0.1& 0.1 \\
\hline
$ \ell^+ \ell^- j j b\bar{b}  $ &$ 1950$&$246$& 44 & $ 3.4 $\\
\hline
\end{tabular}
\end{center}
  \caption{Effects of the kinematical cuts on the production cross section (fb) at the 14 TeV LHC for the $\ell^+\ell^-  jjb\bar{b}$ signal ($M_X=200$ GeV) and backgrounds.}\label{table-effect3}
\end{table}

\begin{figure}
\begin{center}
\includegraphics[width=0.45\textwidth]{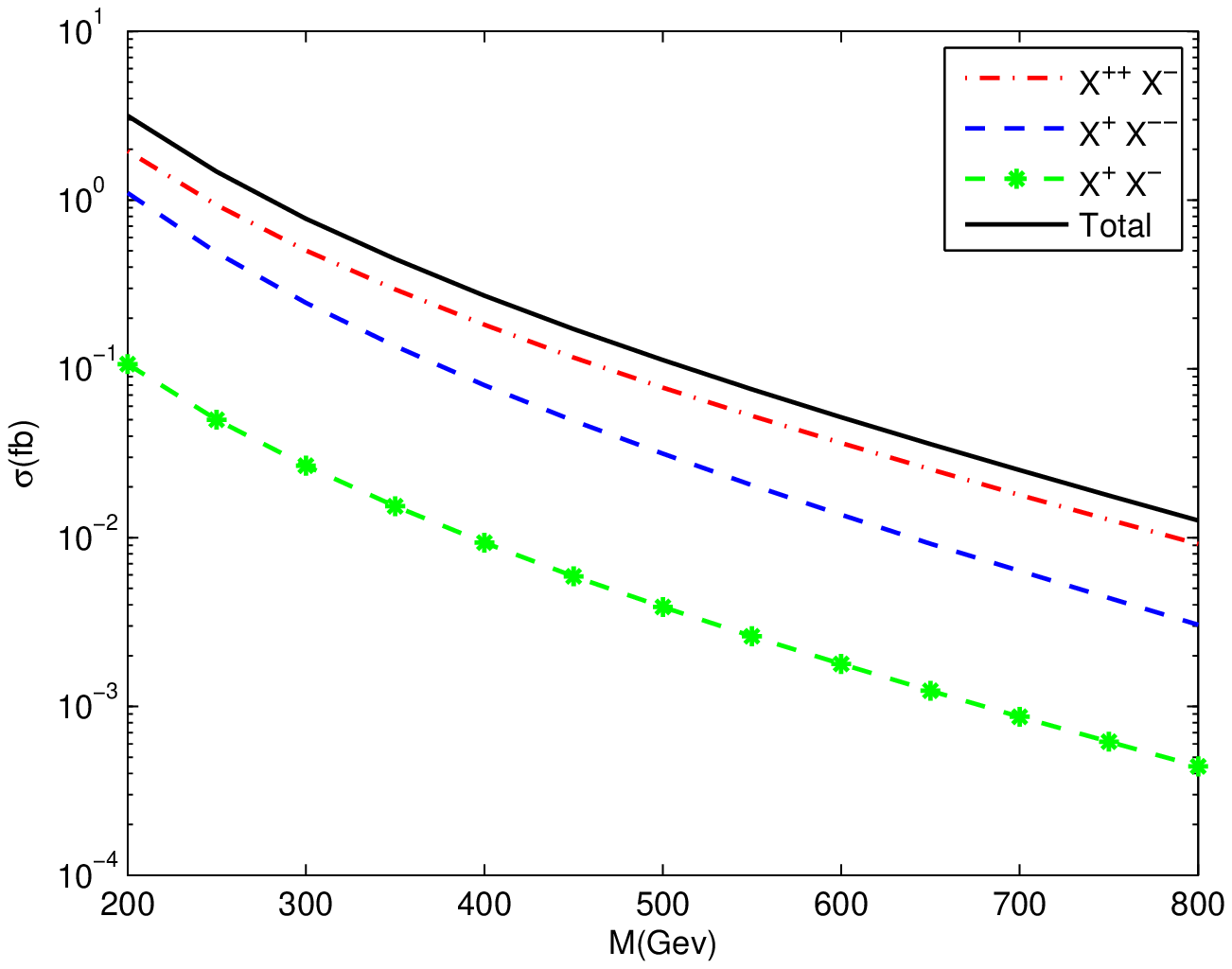}
\includegraphics[width=0.45\textwidth]{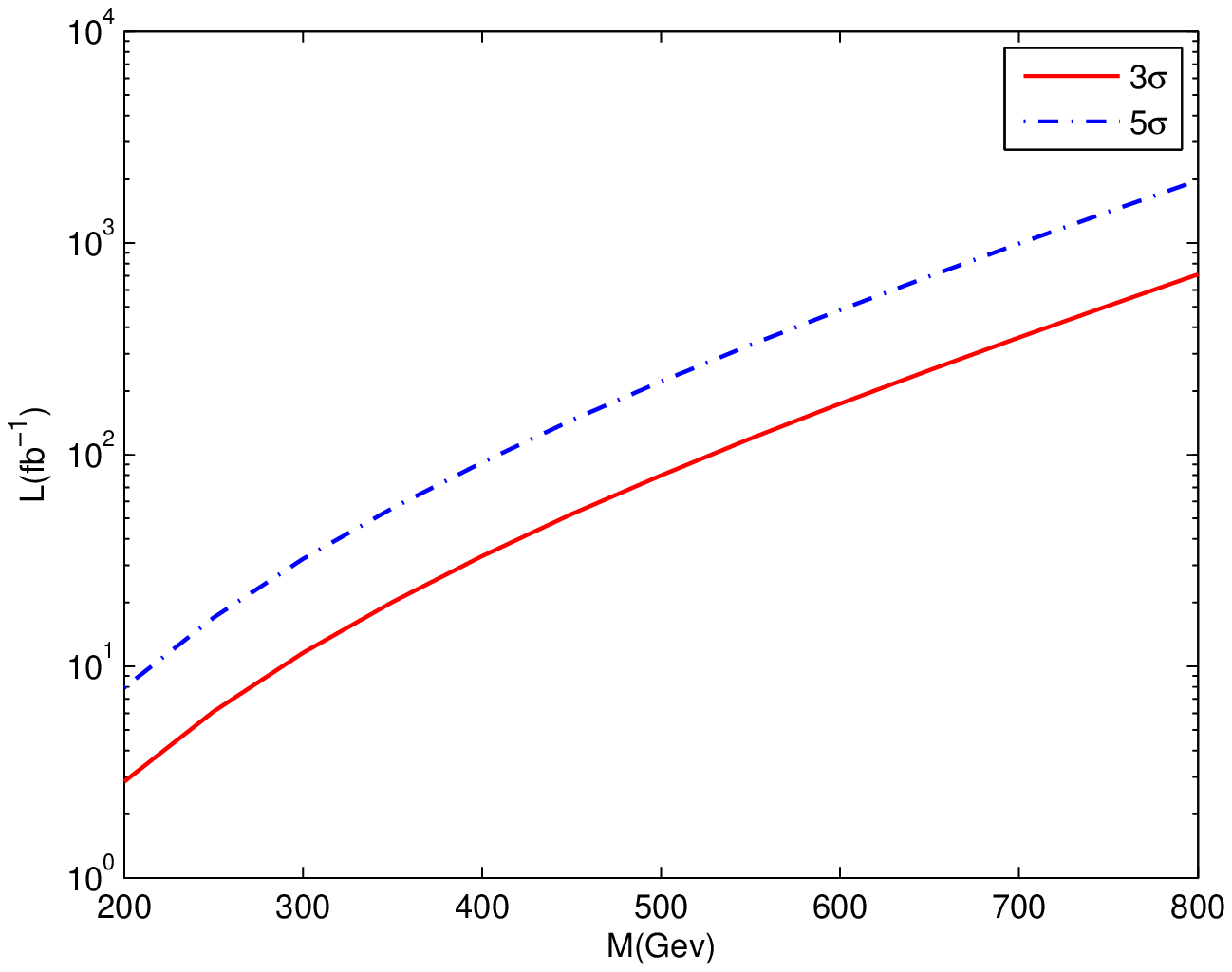}\\
\caption{Left: final lepton triplet signal production rate for $\ell^+\ell^-  jj Z(b\bar{b} )$ channel at 14 TeV LHC. Right: needed luminosity to observe different mass triplet leptons via $\ell^+\ell^-  jj Z(b\bar{b} )$ channel for 3 $\sigma$ and 5 $\sigma$ significances at the 14 TeV LHC.}\label{figure-cs3}
\end{center}
\end{figure}

Once again, we can reconstruct the invariant mass $M(\ell b\bar{b})$ or  $M(\ell jj)$ as an exotic lepton resonance for signal, and the background is very suppressed if we only consider the events under the signal resonances.
After considering the basic acceptance cuts, decay branching fractions of exotic leptons and a b-tagging efficiency of 70\%, the final signal cross section of  $\ell^+\ell^-  jjb\bar{b}$ is similar as $\ell^+\ell^-  jj Z(\ell^+\ell^- )$ channel. The cross section for various $X$ masses is given in the Fig.~\ref{figure-cs3}.
The needed integrated luminosity to observe different mass triplet leptons is plotted in the right panel of Fig.\ref{figure-cs3}: the values are very similar to the leptonic $Z$ channel in Fig.~\ref{figure-lum2} and it would be interesting to improve the sensitivity by combining the two channels which are only distinguished by the well-known $Z$ decays.

\begin{figure}
\begin{center}
\includegraphics[width=0.45\textwidth]{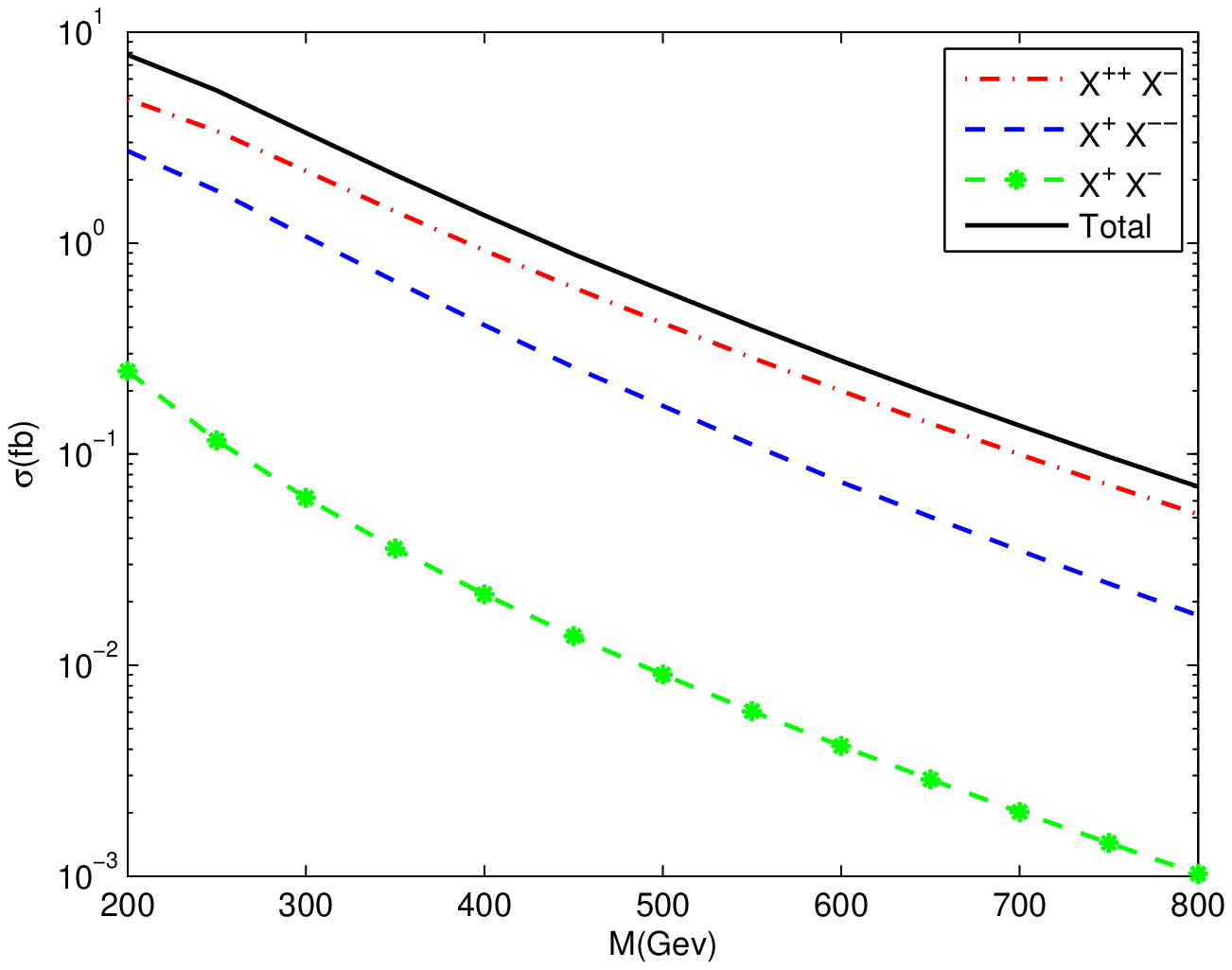}
\includegraphics[width=0.45\textwidth]{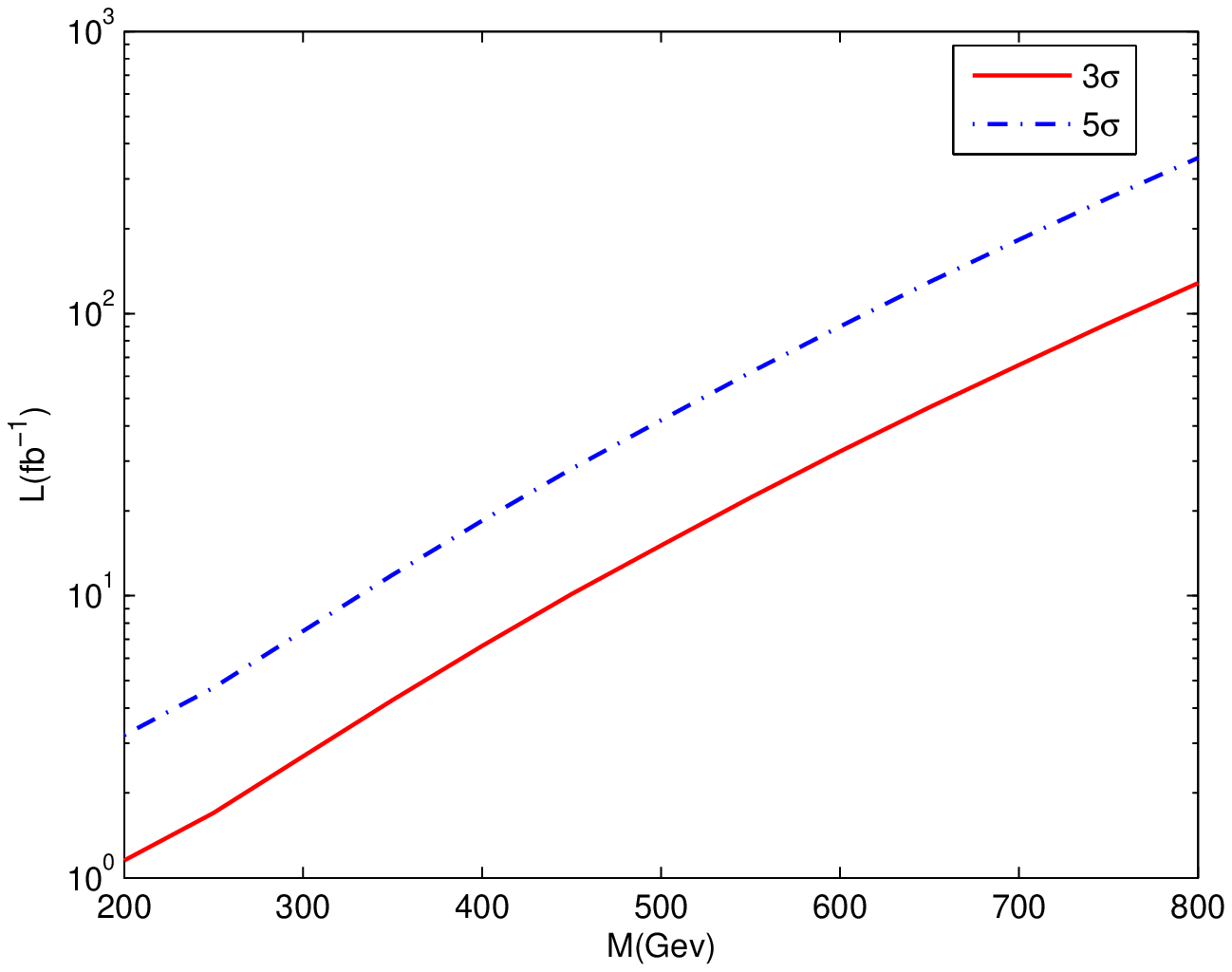}\\
\caption{Left: final lepton triplet signal production rate for $\ell^+\ell^-  jj H(b\bar{b} )$ channel at 14 TeV LHC. Right: needed luminosity to observe different mass triplet leptons via $\ell^+\ell^-  jj H(b\bar{b} )$ channel for 3 $\sigma$ and 5 $\sigma$ significance at the 14 TeV LHC.}\label{figure-cs4}
\end{center}
\end{figure}

The singly charged lepton $X^{\pm}$ has considerable branching fraction of its decay to the Higgs boson (125 GeV), as shown in Fig.\ref{figure-TripletXdecaybranch}.
The Higgs decay to $b\bar{b}$ is dominant, so we can repeat the above analysis of the signal and background of $\ell^+\ell^-  jj H(b\bar{b} )$ by simply changing the $b \bar{b}$ invariant mass cut around the Higgs mass. Here, we just give the final signal cross section of  $\ell^+\ell^-  jj H(b\bar{b})$ and the needed integrated luminosity in Fig.\ref{figure-cs4}.
The final cross section is larger and needed luminosity is less than the $Z(b\bar{b})$ mode because of the largest Higgs decay branching fraction to $b\bar{b}$ pair.

\begin{figure}
\begin{center}
\includegraphics[width=0.6\textwidth]{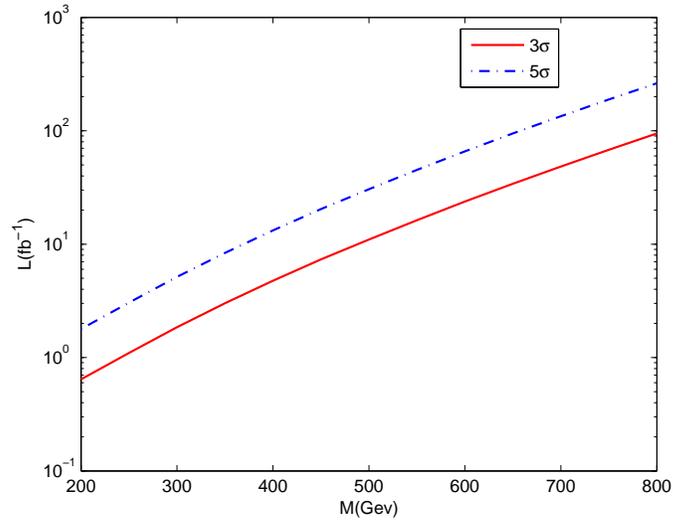}\\
\caption{Needed luminosity to observe different mass triplet leptons via combined $X^{++}X^{-}/X^{+}X^{--}$ channel for 3 $\sigma$ and 5 $\sigma$ significance at the 14 TeV LHC.}\label{figure-lum5}
\end{center}
\end{figure}

In conclusion, we have shown three possible detection modes for $X^{++}X^{-}/X^{+}X^{--}$ associated productions, not considered before, and all the backgrounds are very small after kinematic cuts and resonance reconstruction. We can combine the signals of the three channels together to reduce the integrated luminosity needed to discover or exclude the exotic lepton triplet.
After signature combination, it is possible to find the lepton triplet in the $X^{++}X^{-}/X^{+}X^{--}$ channel if its mass is not larger than 500 GeV as shown in the Fig.\ref{figure-lum5}.

\subsection{Singly charged lepton detection via $X^{+}X^{-}$ pair production}

Although the $X^{+}X^{-}$ pair production can contribute the same signal as the associated production to detect the exotic singly charged leptons, this channel also provides special final states which include two leptons and two neutral bosons. The $X^{+}X^{-}$ pair can decay to two $Z$ bosons, two Higgs bosons or one $Z$ boson and one Higgs boson. All the neutral bosons are possible to be detected via $b\bar{b}$ or leptonic decay. We first consider following processes
\begin{eqnarray}
pp & \rightarrow & X^{+}X^{-} \rightarrow \ell^+ Z\ell^- Z  \rightarrow \ell^+ \ell^- b\bar{b}b\bar{b} \\
%\end{equation}
%
%\begin{equation}
pp&\rightarrow& X^{+}X^{-} \rightarrow \ell^+ H\ell^- H  \rightarrow \ell^+ \ell^- b\bar{b}b\bar{b} \\
%\end{equation}
%
%\begin{equation}
pp& \rightarrow& X^{+}X^{-} \rightarrow \ell^+ Z\ell^- H  \rightarrow \ell^+ \ell^- b\bar{b}b\bar{b}
\end{eqnarray}

The signal final state contains two $b\bar{b}$ pairs from $Z$ or Higgs boson decays with two leptons and no missing energy. Unfortunately, the $X^{+}X^{-}$ pair production rate is small for lepton triplet model. The needed luminosity for LHC to observe the triplet leptons via this channel is rather large without any advantage compared to other channels.

\subsection{Lepton flavor violating signature of the triplet model at LHC }

So far, we have considered channels where at least one gauge boson, either a $W$ or a $Z$, from the exotic leptons decays leptonically thus paying a high price in terms of branching ratios.
Being able to use final states where all gauge bosons decay hadronically would allow us to gain in signal strength considerably, also taking into account that all exotic leptons, except the neutral one, have the decay mode $l jj$.
The signal, however, would be overwhelmed by $l^+ l^- jjjj$ events generated by a $Z$ or photon s-channel.
One way to escape from the background is to consider events with lepton flavor violation, i.e. events with one electron and one muon.
Such events cannot be generated by a $Z$ or photon, therefore the SM background will be dominated by $W^+ W^- jj jj$ events.
One possible way to distinguish signal over background is the absence of invisible particles in the signal, therefore a veto on missing energy may reduce the background to reasonable levels.

There is no reason to believe that exotic leptons only mix with one generation of SM leptons, and in the following we will work in the most favourable situation where the mixing parameters with three generations are the same, therefore a half of the events analyzed before have leptons of different flavor.
All the exotic leptons pair production and associated production processes can provide the lepton flavor changing signature:
\begin{equation}
pp\rightarrow X^{--}X^{++} \rightarrow e^- W^{-}\mu^+W^{+} +h.c. \rightarrow e^- \mu^+ j j jj+h.c.
  \label{31}
\end{equation}
\begin{equation}
pp\rightarrow X^{--}X^{+}+h.c. \rightarrow  e^-  W^{-}\mu^+Z(H) + h.c.  \rightarrow e^- \mu^+ j j jj+h.c.
  \label{32}
\end{equation}
\begin{equation}
pp\rightarrow X^{-}X^{+} \rightarrow e^-\mu^+ZZ(ZH,HH) +h.c. \rightarrow e^- \mu^+ j j jj+h.c.
  \label{33}
\end{equation}
In these signal processes, only the hadronic decay mode of the gauge or Higgs boson is considered in order to avoid confusion with the lepton part.
Here, we do not need b-tagging on the jets from Higgs decay, although the light Higgs (125 GeV) boson decay dominantly into $b\bar{b}$.

The precise rate of this signal is sensitive to the coupling of the exotic leptons to electrons and muons, and the results can be generalised to a generic situation by rescaling the signal events by a factor
\begin{equation}
\frac{9}{4} (1-\zeta_\tau)^2 \frac{4 |v_e|^2 |v_\mu|^2}{(|v_e|^2 + |v_\mu|^2)^2} < \frac{9}{4}\,.
\end{equation}

\begin{figure}
\begin{center}
\includegraphics[width=0.45\textwidth]{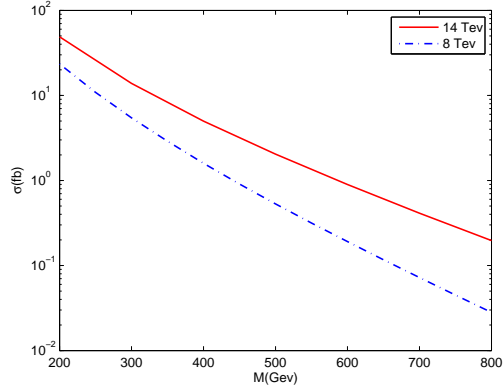}\\
\caption{The final cross sections for the lepton flavor violating signature at 8TeV and 14TeV LHC}\label{figure5555-1}
\end{center}
\end{figure}

\begin{figure}
\begin{center}
\includegraphics[width=0.45\textwidth]{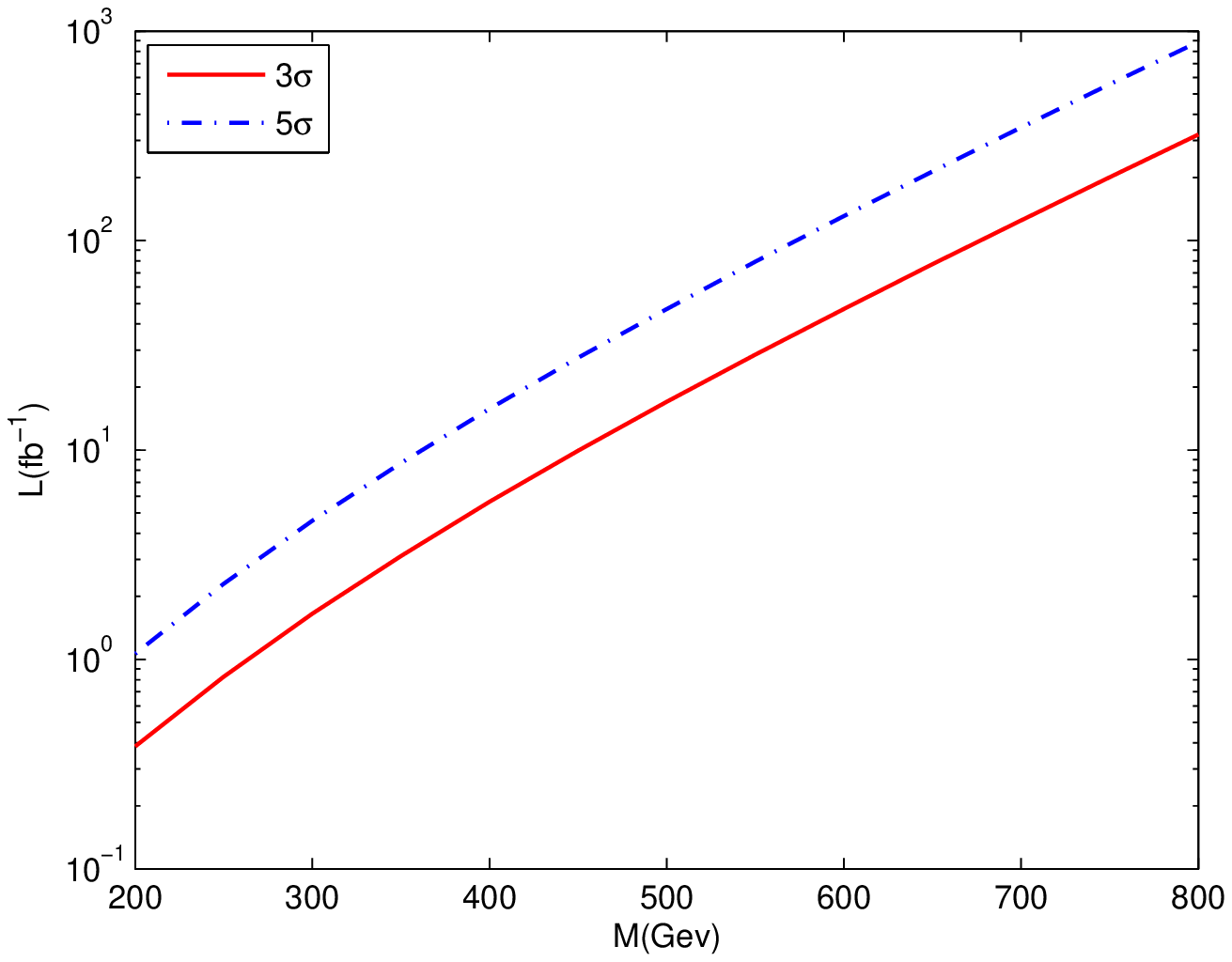}
\includegraphics[width=0.45\textwidth]{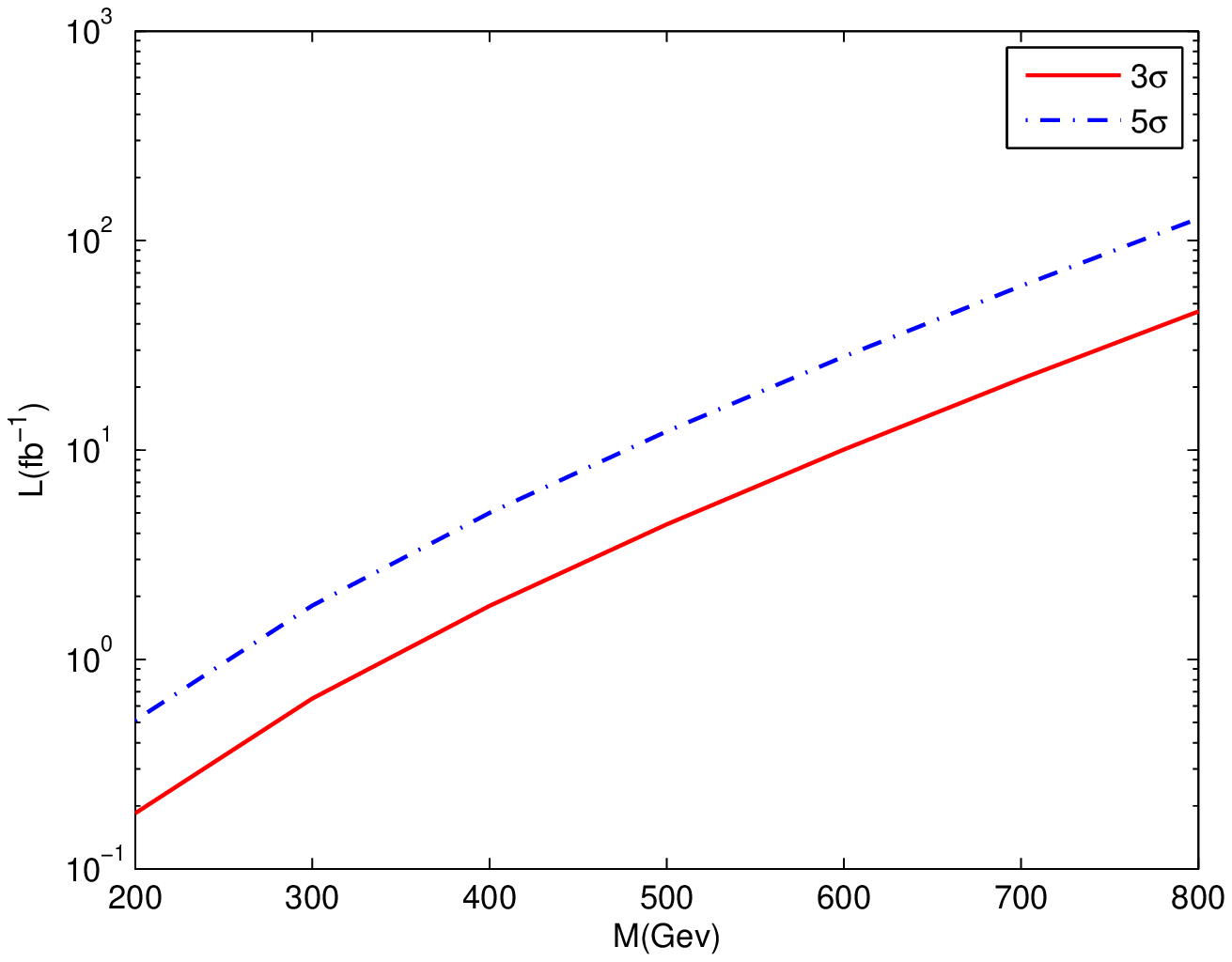}\\
\caption{Needed luminosity to observe different mass triplet leptons via lepton flavor violating processes at the 8 TeV (left) and 14 TeV (right) LHC.}\label{figure5555-2}
\end{center}
\end{figure}

The main background is generated by $W$ pair production with jets, therefore we will consider in the following processes giving
\begin{equation}
pp\rightarrow e^+\mu^-\nu_e\bar{\nu_\mu}+ hadrons\,.
\end{equation}
Even after requiring the missing energy veto cut $ E\!\!\!\slash_T < 25 {\,\rm GeV}$, the remaining background cross section is still larger than $10$ pb, thus we need to add more kinematical cuts to extract the signal.
In the following we require the presence of exactly 4 jets, and the basic kinematic acceptance cuts on the
transverse momentum, rapidity, missing transverse energy, and particle separation with the same values as in the above sections. We impose the $W/Z/H$ mass cut on the jets as before to ensure that they come from gauge or Higgs boson decays.
As there are four jets in the final states, all the pair combinations should be taken into account. If there is a combination in which the two jet pairs invariant masses are both in the W/Z or Higgs boson mass range, then the event passes the $W/Z/H$ mass cut. Then we further impose the mass matching cut on the events
\begin{equation}
| M(jje) - M( \mu jj ) | < 20 \,\rm GeV.
\end{equation}
The two mass cuts do not affect the signal, but can suppress the background to manageable levels. The final cross sections for the lepton flavor violating signature at 8TeV and 14TeV LHC are given in Fig.~\ref{figure5555-1}.
The cross sections are large due to the large branching ratios of $W/Z/H$ boson hadronic decay. We also see in Fig.~\ref{figure5555-2} that an analysis of the 20/fb data at 8 TeV have the potential to probe the lepton triplet model for small masses ($M_X<400$GeV) for our choice of the flavor structure of the couplings.
In the right panel of Fig.~\ref{figure5555-2} we also show the needed luminosity for discovery/exclusion at 14 TeV, showing that the potential to observe the lepton flavor violating channel can be better than the former optimal $\ell^- \ell^+ jj \ell^-\bar{\nu}(\ell^+\nu)$ channel.

\section{Conclusions}

In this work we considered the phenomenology at the LHC of a lepton triplet which mixes with ordinary leptons via Yukawa couplings with the SM Higgs doublet.
The best channel to search for them is a tri-lepton final state, where the two same-sign leptons derive from the decays of the doubly charged exotic lepton.
The reach of this search at LHC at 14 TeV has been studied in~\cite{Delgado:2011iz}: in the present work we reconsidered the proposed cuts, including the contribution of associated production of the doubly charged lepton with the singly charged one.
The extra contribution, which is always present, improves the reach.
Finally, we considered other searches based on the decays of the singly charged lepton: even though their sensitivity is lower compared to the same-sign lepton channel, they are essential to determine the nature of the exotic lepton as they are crucially sensitive to the decays of singly-charged lepton, thus allowing to reconstruct the lepton multiplet.
Finally we point out that, in the case where significant mixing to both electrons and muons is present, a search based on two different flavor leptons plus jets without missing energy is very effective, and may allow to pose significant bounds on the exotic lepton mass already with the data at 8 TeV.

{\bf Acknowledgment:} The work of B.~Z. is supported by the
National Science Foundation of China under Grant No. 11075086 and
11135003. G.~C. would like to thank the Physics Department in Tsinghua University for the hospitality during the initial stages of this work, the visit being sponsored by a project of the France-China Particle Physics Laboratory (FCPPL) and an exchange program of the CNRS/NSFC.


\begin{thebibliography}{99}


%\cite{Schmaltz:2005ky}
\bibitem{Schmaltz:2005ky}
  M.~Schmaltz and D.~Tucker-Smith,
  %``Little Higgs review,''
  Ann.\ Rev.\ Nucl.\ Part.\ Sci.\  {\bf 55} (2005) 229
  [hep-ph/0502182].
  %%CITATION = HEP-PH/0502182;%%
  %387 citations counted in INSPIRE as of 29 Aug 2013


%\cite{Cacciapaglia:2009cv}
\bibitem{Cacciapaglia:2009cv}
  G.~Cacciapaglia, A.~Deandrea, S.~R.~Choudhury and N.~Gaur,
  %``T-parity odd heavy leptons at LHC,''
  Phys.\ Rev.\ D {\bf 81} (2010) 075005
  [arXiv:0911.4632 [hep-ph]].
  %%CITATION = ARXIV:0911.4632;%%
  %5 citations counted in INSPIRE as of 29 Aug 2013

%\cite{Frandsen:2009fs}
\bibitem{Frandsen:2009fs}
  M.~T.~Frandsen, I.~Masina and F.~Sannino,
  %``Fourth Lepton Family is Natural in Technicolor,''
  Phys.\ Rev.\ D {\bf 81} (2010) 035010
  [arXiv:0905.1331 [hep-ph]].
  %%CITATION = ARXIV:0905.1331;%%
  %28 citations counted in INSPIRE as of 29 Aug 2013

%\cite{Franceschini:2008pz}
\bibitem{Franceschini:2008pz}
  R.~Franceschini, T.~Hambye and A.~Strumia,
  %``Type-III see-saw at LHC,''
  Phys.\ Rev.\ D {\bf 78} (2008) 033002
  [arXiv:0805.1613 [hep-ph]].
  %%CITATION = ARXIV:0805.1613;%%
  %98 citations counted in INSPIRE as of 29 Aug 2013

%\cite{delAguila:2008cj}
\bibitem{delAguila:2008cj}
  F.~del Aguila and J.~A.~Aguilar-Saavedra,
  %``Distinguishing seesaw models at LHC with multi-lepton signals,''
  Nucl.\ Phys.\ B {\bf 813}, 22 (2009)
  [arXiv:0808.2468 [hep-ph]].
  %%CITATION = ARXIV:0808.2468;%%
  %190 citations counted in INSPIRE as of 24 Nov 2013

%\cite{delAguila:2008hw}
\bibitem{delAguila:2008hw}
  F.~del Aguila and J.~A.~Aguilar-Saavedra,
  %``Electroweak scale seesaw and heavy Dirac neutrino signals at LHC,''
  Phys.\ Lett.\ B {\bf 672}, 158 (2009)
  [arXiv:0809.2096 [hep-ph]].
  %%CITATION = ARXIV:0809.2096;%%
  %72 citations counted in INSPIRE as of 24 Nov 2013

%\cite{delAguila:2008pw}
\bibitem{delAguila:2008pw}
  F.~del Aguila, J.~de Blas and M.~Perez-Victoria,
  %``Effects of new leptons in Electroweak Precision Data,''
  Phys.\ Rev.\ D {\bf 78}, 013010 (2008)
  [arXiv:0803.4008 [hep-ph]].
  %%CITATION = ARXIV:0803.4008;%%
  %82 citations counted in INSPIRE as of 24 Nov 2013

%\cite{Arhrib:2009mz}
\bibitem{Arhrib:2009mz}
  A.~Arhrib, B.~Bajc, D.~K.~Ghosh, T.~Han, G.~-Y.~Huang, I.~Puljak and G.~Senjanovic,
  %``Collider Signatures for Heavy Lepton Triplet in Type I+III Seesaw,''
  Phys.\ Rev.\ D {\bf 82}, 053004 (2010)
  [arXiv:0904.2390 [hep-ph]].
  %%CITATION = ARXIV:0904.2390;%%
  %55 citations counted in INSPIRE as of 24 Nov 2013



%\cite{delAguila:2000aa}
\bibitem{delAguila:2000aa}
  F.~del Aguila, M.~Perez-Victoria and J.~Santiago,
  %``Effective description of quark mixing,''
  Phys.\ Lett.\ B {\bf 492}, 98 (2000)
  [hep-ph/0007160].
  %%CITATION = HEP-PH/0007160;%%
  %76 citations counted in INSPIRE as of 24 Nov 2013

%\cite{delAguila:2000rc}
\bibitem{delAguila:2000rc}
  F.~del Aguila, M.~Perez-Victoria and J.~Santiago,
  %``Observable contributions of new exotic quarks to quark mixing,''
  JHEP {\bf 0009}, 011 (2000)
  [hep-ph/0007316].
  %%CITATION = HEP-PH/0007316;%%
  %94 citations counted in INSPIRE as of 24 Nov 2013

%\cite{AguilarSaavedra:2009es}
\bibitem{AguilarSaavedra:2009es}
  J.~A.~Aguilar-Saavedra,
  %``Identifying top partners at LHC,''
  JHEP {\bf 0911} (2009) 030
  [arXiv:0907.3155 [hep-ph]].
  %%CITATION = ARXIV:0907.3155;%%
  %77 citations counted in INSPIRE as of 29 Aug 2013


%\cite{Cacciapaglia:2010vn}
\bibitem{Cacciapaglia:2010vn}
  G.~Cacciapaglia, A.~Deandrea, D.~Harada and Y.~Okada,
  %``Bounds and Decays of New Heavy Vector-like Top Partners,''
  JHEP {\bf 1011} (2010) 159
  [arXiv:1007.2933 [hep-ph]].
  %%CITATION = ARXIV:1007.2933;%%
  %20 citations counted in INSPIRE as of 29 Aug 2013

%\cite{Buchkremer:2013bha}
\bibitem{Buchkremer:2013bha}
  M.~Buchkremer, G.~Cacciapaglia, A.~Deandrea and L.~Panizzi,
  %``Model Independent Framework for Searches of Top Partners,''
  arXiv:1305.4172 [hep-ph].
  %%CITATION = ARXIV:1305.4172;%%
  %5 citations counted in INSPIRE as of 29 Aug 2013


%\cite{Aguilar-Saavedra:2013qpa}
\bibitem{Aguilar-Saavedra:2013qpa}
  J.~A.~Aguilar-Saavedra, R.~Benbrik, S.~Heinemeyer and M.~Perez-Victoria,
  %``A handbook of vector-like quarks: mixing and single production,''
  Phys.\ Rev.\ D {\bf 88}, 094010 (2013)
  [arXiv:1306.0572 [hep-ph]].
  %%CITATION = ARXIV:1306.0572;%%
  %18 citations counted in INSPIRE as of 24 Nov 2013



%\cite{DeSimone:2012fs}
\bibitem{DeSimone:2012fs}
  A.~De Simone, O.~Matsedonskyi, R.~Rattazzi and A.~Wulzer,
  %``A First Top Partner's Hunter Guide,''
  JHEP {\bf 1304} (2013) 004
  [arXiv:1211.5663 [hep-ph]].
  %%CITATION = ARXIV:1211.5663;%%
  %16 citations counted in INSPIRE as of 29 Aug 2013



%\cite{Cirelli:2005uq}
\bibitem{Cirelli:2005uq}
  M.~Cirelli, N.~Fornengo and A.~Strumia,
  %``Minimal dark matter,''
  Nucl.\ Phys.\ B {\bf 753} (2006) 178
  [hep-ph/0512090].
  %%CITATION = HEP-PH/0512090;%%
  %230 citations counted in INSPIRE as of 29 Aug 2013

%\cite{Delgado:2011iz}
\bibitem{Delgado:2011iz}
  A.~Delgado, C.~Garcia Cely, T.~Han and Z.~Wang,
  %``Phenomenology of a lepton triplet,''
  Phys.\ Rev.\ D {\bf 84} (2011) 073007
  [arXiv:1105.5417 [hep-ph]].
  %%CITATION = ARXIV:1105.5417;%%
  %7 citations counted in INSPIRE as of 15 Aug 2013

%\cite{Alloul:2013raa}
\bibitem{Alloul:2013raa}
  A.~Alloul, M.~Frank, B.~Fuks and M.~R.~de Traubenberg,
  %``Doubly-charged particles at the Large Hadron Collider,''
  arXiv:1307.1711 [hep-ph].
  %%CITATION = ARXIV:1307.1711;%%
  %3 citations counted in INSPIRE as of 01 Oct 2013


%\cite{Cacciapaglia:2011fx}
\bibitem{Cacciapaglia:2011fx}
  G.~Cacciapaglia, A.~Deandrea, L.~Panizzi, N.~Gaur, D.~Harada and Y.~Okada,
  %``Heavy Vector-like Top Partners at the LHC and flavour constraints,''
  JHEP {\bf 1203} (2012) 070
  [arXiv:1108.6329 [hep-ph]].
  %%CITATION = ARXIV:1108.6329;%%
  %15 citations counted in INSPIRE as of 26 Aug 2013

  %\cite{ALEPH:2005ab}
\bibitem{LEP}
  S.~Schael {\it et al.}  [ALEPH and DELPHI and L3 and OPAL and SLD and LEP Electroweak Working Group and SLD Electroweak Group and SLD Heavy Flavour Group Collaborations],
  %``Precision electroweak measurements on the $Z$ resonance,''
  Phys.\ Rept.\  {\bf 427} (2006) 257
  [\href{http://arxiv.org/abs/hep-ex/0509008}{hep-ex/0509008}].
  %%CITATION = HEP-EX/0509008;%%

%\cite{Feng:1999fu}
\bibitem{Feng:1999fu}
  J.~L.~Feng, T.~Moroi, L.~Randall, M.~Strassler and S.~-f.~Su,
  %``Discovering supersymmetry at the Tevatron in wino LSP scenarios,''
  Phys.\ Rev.\ Lett.\  {\bf 83} (1999) 1731
  [hep-ph/9904250].
  %%CITATION = HEP-PH/9904250;%%
  %152 citations counted in INSPIRE as of 27 Aug 2013


\bibitem{ATLAS}
ATLAS collaboration, ATLAS-CONF-2013-070.

\bibitem{CMS}
CMS collaboration, CMS-PAS-SUS-13-002.

\bibitem{MadGraph}
J. Alwall et al.,
%¡°MadGraph 5: Going Beyond¡±,
JHEP 1106, 128,
arXiv:1106.0522 [hep-ph] (2011).

%\cite{Ball:2007zza}
\bibitem{Ball:2007zza}
  G.~L.~Bayatian {\it et al.}  [CMS Collaboration],
  %``CMS technical design report, volume II: Physics performance,''
  J.\ Phys.\ G {\bf 34}, 995 (2007).



\end{thebibliography}
\end{document}